\documentstyle[multicol,eqsecnum,aps,psfig,epsf]{revtex}
\newcommand{\be}{\begin{equation}}
\newcommand{\ee}{\end{equation}}
\newcommand{\bea}{\begin{eqnarray}}
\newcommand{\eea}{\end{eqnarray}}

\begin{document}
\draft
\title{Intermittency exponents and energy spectrum of the Burgers and
KPZ equations with correlated noise}
\author{Mahendra K. Verma
\thanks{email:mkv@iitk.ac.in}}
\address{Department of
Physics, Indian Institute of Technology, \\
Kanpur -- 208016, INDIA}
\date{\today}
\maketitle

\begin{abstract}
We numerically calculate the energy spectrum, intermittency exponents,
and probability density  $P(u')$ of the one-dimensional
Burgers and KPZ equations with correlated noise.  We have used
pseudo-spectral method for our analysis.  When $\sigma$ of the  noise
variance of the Burgers equation (variance $\propto k^{-2 \sigma}$) 
exceeds 3/2, large shocks appear in the velocity profile leading to
$<|u(k)|^2> \propto k^{-2}$, and structure function
$<|u(x+r,t)-u(x,t)|^q> \propto r$ suggesting that the Burgers equation
is intermittent for this range of $\sigma$.  For $-1 \le \sigma \le
0$, the profile is dominated by noise, and the spectrum $<|h(k)|^{2}>$
of the corresponding KPZ equation is in close agreement with Medina et
al.'s renormalization group predictions.  In the intermediate range $0
< \sigma <3/2$, both noise and well-developed shocks are seen,
consequently the exponents slowly vary from RG regime to a
shock-dominated regime.  The probability density $P(h)$ and $P(u)$ are
gaussian for all $\sigma$, while $P(u')$ is gaussian for $\sigma=-1$,
but steadily becomes nongaussian for larger $\sigma$; for negative
$u'$, $P(u') \propto \exp(-a x)$ for $\sigma=0$, and approximately
$\propto u'^{-5/2}$ for $\sigma > 1/2$.  We have also calculated the
energy cascade rates for all $\sigma$ and found a constant flux for
all $\sigma \ge 1/2$.

\end{abstract}
\vspace{0.5in} 

\section{Introduction}

Stochastic Burgers equation has been studied extensively because of
its close connection with the Navier-Stokes equation.  Recently it has
been found that Burgers equation with correlated noise shows
intermittency \cite{Chek1,Poly,Hayo}, however, these calculations have
been done for specific degrees of noise correlation. In this paper we
vary the applied noise from uncorrelated regime to strongly correlated
regime and study the energy spectrum of $h$ and $u$, as well as
intermittency exponents and the probability densities of $u$ and
$du/dx$, where $u(x,t)$ is the velocity function appearing in the
Burgers equation, and $h(x,t)$ is the surface height of KPZ equation.

The one-dimensional Burgers equation \cite{Burg} is
\begin{equation}
\label{Burgers}
\frac{\partial u}{\partial t}+ \lambda u \frac{\partial u}{\partial x}
= \nu \frac{\partial^2 u}{\partial{x^2}} +\zeta,
\end{equation}
where $u$ is the velocity field, $\lambda$ is the strength of the
nonlinear term ($\lambda=1$ in the standard equation), $\nu$ is the
viscosity, and $\zeta$ is the noise.  We assume that the noise
$\zeta(k,t)$ is gaussian ($\zeta(k)$ is the Fourier transform of
$\zeta(x)$), and in Fourier space follows distribution
\begin{equation}
\label{Burgnoisek}
<\zeta(k,t)\zeta(k^{\prime },t^{\prime })>=2Dk^{-2\sigma }\left( 2\pi
\right) ^2\delta (k+k^{\prime })\delta (t-t^{\prime })
\end{equation}
 
Tatsumi and Kida \cite{Kida1}, Kida \cite{Kida2}, Gotoh \cite{Goto},
and Bouchaud et al. \cite{Bouc} have solved the noiseless Burgers
equation ($\zeta=0$) exactly in limit of large time $t$ and zero
viscosity. The solution is a linear profile with sharp
discontinuities, which are called shocks. This shock solution yields
structure function $<|u(x+r,t)-u(x,t)|^q> \propto r$ indicating that
the $u$ is intermittent.  The solution also yields the energy spectrum
$<|u(k)|^{2}/2> \propto k^{-2}$ ($u(k)$ is the Fourier transform of
$u(x)$).

The noisy Burgers equation as well as the noisy Kardar-Parisi-Zhang (KPZ)
 equation have been studied by many authors.  The Kardar-Parisi-Zhang
 (KPZ) equation, which describes a generic set of surface growth
 phenomena, is closely related to the Burgers equation. The
 replacement of $u$ by $-\partial h/ \partial x$ in the Burgers
 equation yields KPZ equation:
\begin{equation}
\label{KPZ}
\frac{\partial h}{\partial t}=\frac{\lambda}{2} \left( \frac{\partial
h}{\partial x}\right)^2 + \nu \frac{\partial^2 h}{\partial x^2}
+f({\bf x},t),
\end{equation}
where $h(x,t)$ is the height of the surface profile at position $x$
and at time $t$, $\lambda $ is the strength of the nonlinearity, $\nu
$ is the diffusion coefficient, and $f$ is the forcing function.  We
again assume that the noise $f(k,t)$ is gaussian, and in Fourier space
follows distribution
\begin{equation}
\label{KPZnoisek}
<f(k,t)f(k^{\prime },t^{\prime })>=2Dk^{-2\rho }\left( 2\pi \right)
^2\delta (k+k^{\prime })\delta (t-t^{\prime }).
\end{equation}
It is easy to see that $|u(k)|^{2} = k^2 |h(k)|^2$ and
$\sigma=\rho-1$.  In the following several paragraphs we will describe
the work done by various researchers regarding energy spectrum,
intermittency exponents, and probability densities of $h$ and $u$ when
noise is present in the system.

Eq. (\ref{KPZ}) has been solved by Kardar-Parisi-Zhang \cite{KPZ} for
$\sigma=-1$ and by Medina et al. \cite{Medi} for $-1< \sigma \leq 0$
using renormalization group analysis.  They have calculated the
roughening exponents $\chi_{KPZ}$ and $\beta_{KPZ}$, which
characterize the dynamical properties of the equation, e.g.,
$<|h(k)|^2> \propto k^{-2 \chi_{KPZ} -1}$.  The RG analysis is not
applicable beyond $\sigma =0$.  Using the connection of the Burgers
equation with the KPZ equation, the dynamical exponents of the Burgers
equation is also automatically solved for this set of $\sigma$.
Recently, Chattopadhyay and Bhattacharjee \cite{ChatJKB} have applied
mode coupling scheme and obtained the same exponents for KPZ equation
for $-1 \le \sigma \le 0$.

For $\sigma=-1$, Barabasi and Stanley \cite{Bara} have shown that in
1D KPZ equation, the function $dh/dx (=-u)$ has a gaussian probability
density under steady-state.  Since $h'$ is gaussian, we can
immediately determine the structure function
$T_q(r)=<|h(x+r)-h(x)|^q>$ using the following arguments.  Since
$\Delta h \approx r h'$ for small $r$, $\Delta h [=h(x+r)-h(x)]$ 
will also have a gaussian probability density, i.e., 
\be 
P(\Delta h)=\frac{1}{\sigma_r
\sqrt \pi} \exp\left( -\frac{(\Delta h)^2}{\sigma_r^2} \right) .  
\ee
Therefore, 
\bea <|\Delta h|^q> & = & \int_{-\infty}^{\infty} P(\Delta
h) |\Delta h|^q d\Delta h \nonumber \\ & \propto & \sigma_r^q.  
\eea
In the region where $<|\Delta h|^q>$ is a powerlaw (region of our
interest), $\sigma_r$ will also be a powerlaw, say, $\sigma_r \propto
r^\chi$.  Therefore, $<|\Delta h|^q> \propto r^{\chi q}$.

The above discussion clearly shows that if the derivative of the
function has a gaussian probability density, then the exponent of the
$q-$th order structure function (denoted by $\zeta_q$) is proportional
to $q$.  These kind of functions are called {\em non-intermittent}
functions.  On the other hand, if the probability density of the
derivative is nongaussian, or the exponent of $q-$th order structure
function is not proportional to $q$, then the function is said to be
{\em intermittent}. The intermittent functions have typically powerlaw
tails instead of gaussian tails in their probability densities, hence,
probability of occurrence of large value of the function is higher for
the intermittent functions then the non-intermittent functions.  In
other words, large events, which are absent in non-intermittent
systems, occur in intermittent systems in a bursty manner.  In fluid
turbulence, it has been found that $P(u)$ is gaussian, whereas
$P(u'_L)$ and $P(u'_T)$ (where $L$ and $T$ denote longitudinal and
transverse components) have powerlaw tails for small $r$.  The
exponents of the structure functions are not proportional to $q$
either.  Hence, fluid turbulence is said to have intermittency. These
observations are explained using localized vortices.  For further
discussions on intermittency in fluid turbulence, refer to Frisch
\cite{Frisbook} and references therein.

Chekhlov and Yakhot \cite{Chek1}, and Hayot and Jayaprakash
\cite{Hayo} numerically solved the structure function of the Burgers
equation with correlated noise for a range of $\sigma$.  For
$\sigma=1/2$, Chekhlov and Yakhot \cite{Chek1} obtained $<|\Delta
u(r)|^q> \propto r$ and claimed that the Burgers equation is
intermittent according to the above definitions.  Hayot and Jayprakash
\cite{Hayo} found similar results for $0 \le \sigma \le 1/2$.  This
behaviour was attributed to the shocks present in the system
\cite{Chek1,Hayo}.
 
Regarding the energy spectrum, Chekhlov and Yakhot \cite{Chek1}
obtained Kolmogorov's energy spectrum, i.e., $|u(k)|^{2} \propto
k^{-5/3}$ for $\sigma=1/2$.  They argued the constancy of energy flux
to be the reason for Kolmogorov's spectrum in the noisy Burgers
equation with $\sigma=1/2$.  In this paper we will show that energy
flux is constant for all $\sigma \ge 1/2$, yet the spectral exponent
varies from 5/3 to 2.  Hence, Chekhlov and Yakhot \cite{Chek1} claim
that the constancy of flux implies Kolmogorov's spectrum for Burgers
equation is incorrect.

Hayot and Jayaprakash \cite{Hayo} have numerically calculated the
energy spectrum for $-1 \le \sigma \le 1/2$. For $-1 \le \sigma \le
0$, they observed that at large $k$ the behaviour is that of free
field, and at low $k$ is in agreement with the result derived using RG
treatment. In our simulation we do not find any such crossover.  For
$0 \le \sigma \le 1/2$, Hayot and Jayaprakash argue that the shocks
determine the exponents; their numerical exponent for energy is quite
close to the renormalization group formula even though RG is not
expected to work in this regime.  They also find that the exponent
$\beta_{Burg}=1/2$ for the Burgers equation.

Polyakov \cite{Poly} has applied methods of quantum field theory to
calculate probability density $P(u')$ of the Burgers equation with
$\sigma \ge 3/2$, and showed that \be P(u') = \left\{
\begin{array}{ll} \exp[{-u'^{3}/(3 B_1)}] & \mbox{if $u' \rightarrow
\infty$} \\ u'^{-5/2} & \mbox{if $u' \rightarrow -\infty$} \end{array}
\right.  \ee Clearly, the probability density of $u'$ is nongaussian
and has a power-law tail for negative $u'$. Boldyrev
\cite{Bold1,Bold2} extended Polyakov's method to the range $1/2 \le
\sigma \le 3/2$ and calculated $P(u')$.  He derived that $\chi = 1+2
\sigma /3$ for $1/2 \le \sigma \le 3/4$, and $\chi=3/2$ for $\sigma >
3/4$.  Boldyrev \cite{Bold1} suggests that the above formula may also
be valid for $0 \le \sigma \le 3/2$.

Gotoh and Kraichnan \cite{GotoKrai} found the exponent of $P(u')$ for
the negative $u'$ to be $-3$.  Recently, E and Eijnden \cite{E} and
Kraichnan \cite{KraiBurg} argue the exponent to be $-7/2$.  Gurarie and
Migdal \cite{Gura}, and Balkovsky et al. \cite{Balk} have applied
instanton solutions for solving probability density for positive $u'$
and obtained qualitative agreement with the numerical results reported
by Yakhot and Cheklov \cite{Chek2}.  In this paper we will compare our
findings with the above mentioned results.
  
The roughening exponents for KPZ equation $\chi_{KPZ} $ and
$\beta_{KPZ} $ have also been calculated by Zhang by replica method
\cite{Zhan}, and by Hentschel and Family using scaling arguments
\cite{HentFami}.  They find a close agreement with the results of
Medina et al.  \cite{Medi}. Numerically, Peng et al. \cite{Peng} have
calculated the exponents $\chi_{KPZ}$ and $\beta_{KPZ}$ using
finite-difference method, and Amar et al. \cite{Amar} and Meakin and
Jullien \cite{Meak1,Meak2} have calculated using lattice
simulations. Their results are in qualitative agreement with the RG
predictions.

The energy cascade rate is one of the important quantities of interest
in the statistical theory of turbulence.  In this paper we have
computed the energy flux of the Burgers equation for $-1 \le \sigma
<6$.  We find a constant flux for all $\sigma$ beyond 1/2.  In our
simulation we also varied the values of the parameters
($\nu,\lambda,D$) and analyzed its effects.  We find that there are
interesting crossover from KPZ to Edward Wilkinson (EW) equation
depending on the values of the parameter. These crossover results are
reported in the Appendix A.

The outline of the paper is as follows: We restate the structure
function and $P(u')$ of the noiseless Burgers and KPZ equations in
section 2, and the energy spectrum of these equations in section 3.
From the nature of the noise-noise correlation, we divide the $\sigma$
regime in three parts.  This result is discussed in section 4.
Section 5 and 6 contain the simulation method and results
respectively.  Here we calculate the energy spectrum and structure
functions of the Burgers and KPZ equations for various values of
$\sigma$.  The probability densities of $u'$ and $u$ is also discussed
in this section.  Section 7 contains a brief discussion on the cascade
rate of the noisy Burgers equation.  Section 8 contains discussion and
conclusions.

\section{Structure Function Calculation for noiseless case}

\begin{figure}
\centerline{
        \psfig{figure=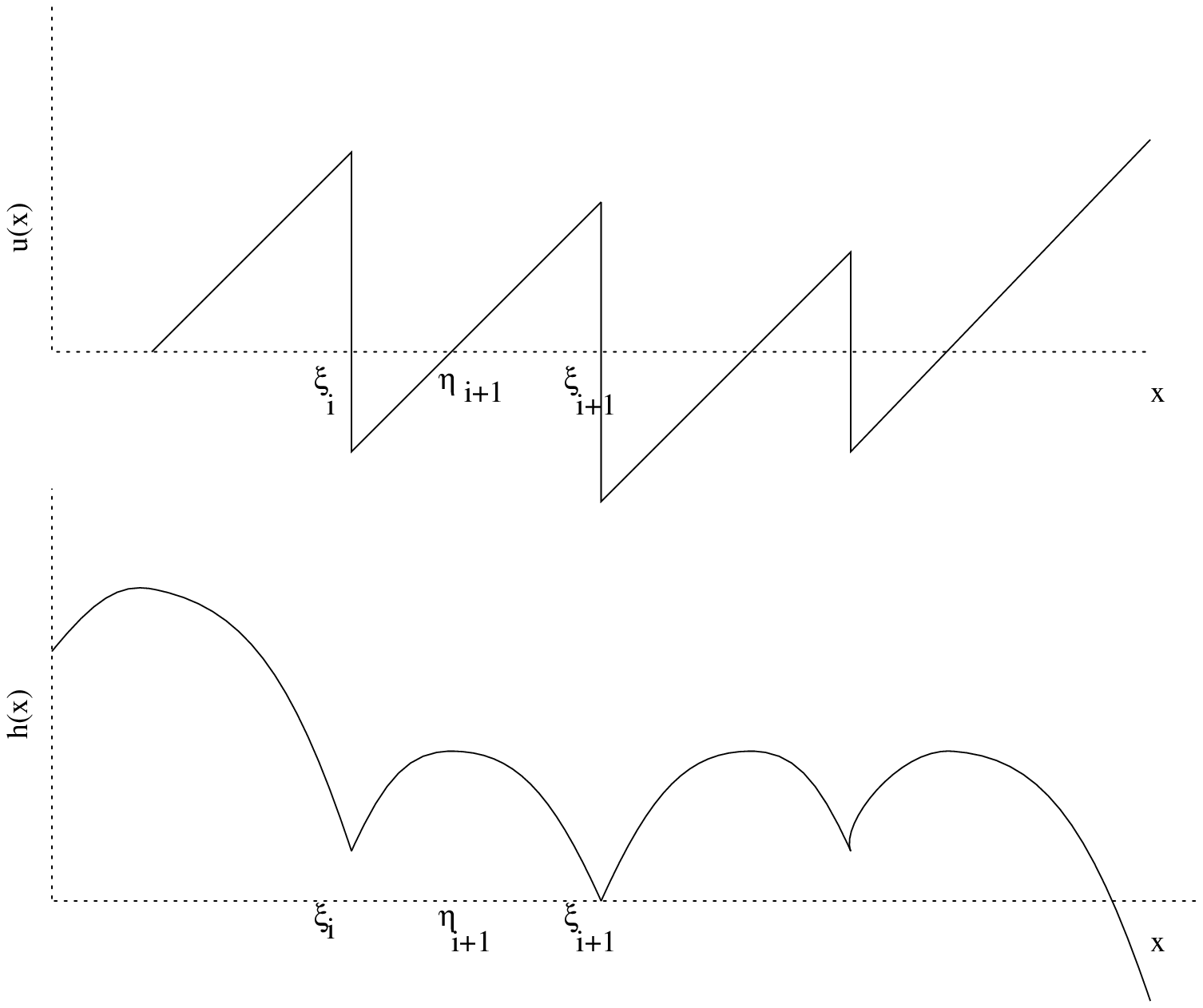,width=12cm,angle=0}}
        \vspace*{0.5cm}
\caption{}
\label{profilefig}
\end{figure}

We will show later that the Burgers equation with strongly correlated
noise has behaviour similar to the noiseless Burgers equation.
Therefore, we will discuss the noiseless equation briefly before
embarking on the noisy Burgers equation.  Burgers \cite{Burg} solved
the Burgers equation exactly at large time $t$ for vanishing
viscosity. The velocity profile at a given time is linear except for
the sharp discontinuities at the shock positions (see
Fig.~\ref{profilefig}).  The structure function of the noiseless
Burgers equation $S_q(r)$ has been calculated using this solution (see
Gotoh \cite{Goto}, Bouchaud et al. \cite{Bouc} and others).  Assuming
the ensemble average to be the same as the spatial average, Gotoh
obtained
\begin{eqnarray}
\label{struct_Burg}
  S_{q}(r) & = & \frac{1}{L} \int_{0}^{L} |u(x+r)-u(x)|^{q} dx
     \nonumber \\ & = & \frac{1}{L} \sum_{i}
     \int_{\xi_{i}}^{\xi_{i+1}-r} \left( \frac{r}{t} \right)^{q} dx
     +\frac{1}{L} \int_{\xi_{i+1}-r}^{\xi_{i+1}} \left( \mu_{i+1} -
     \frac{r}{t}\right)^{q} dx \nonumber \\ & \approx & \left(
     \frac{r}{t} \right)^{q} + \frac{r}{L} \sum_{i} \left( \mu_{i}-
	\frac{r}{t} \right)^{q}
\end{eqnarray}
where $L$ is the length of the box, $\mu_i$ is the shock strength
(defined as the velocity difference across the shock) of the $i$th
shock, and $\xi_{i}$ is the position of the $i$th shock.  The second
term of the equation is due to the discontinuities at the shocks. 
For small $r$ and 
$q>1$, the second term will dominate the first one. Hence,  
for small $r$, $S_q(r)\propto r^{\zeta^{Burg}_q} $ with 
\be
\label{zeta_Burg}
\zeta^{Burg}_q = 1.  
\ee 
Note that this relationship is valid for $
\delta \ll r \ll \Delta$, where $\delta$ is the shock width, and
$\Delta$ is the average distance between two shocks.  For $\nu
\rightarrow 0$, $\delta$ is finite but small (for details, see Saffman
\cite{Saff}).  Since $u$ is continuous within the shock, $S_q(r)=r^q$
for $r \ll \delta$.  For large $r$, $S_q(r)$ is not proportional
to $r$ because of  $(\mu_{i}-r/t)^{q}$ term of Eq.~(\ref{struct_Burg}).
 
A point is in order here.  Tatsumi and Kida \cite{Kida1} showed that
the number of shock fronts decrease with $t$ as $t^{-\gamma }$, where
$0\leq \gamma <1$. Hence in the asymptotic state, there will be only
several shocks, and the distance between the shock fronts will be of
the order of box size. Therefore, asymptotically $\Delta$ will
of the order of the box size $L$, and  $r$ can be comparable to $L$.

\begin{figure}
\centerline{
        \psfig{figure=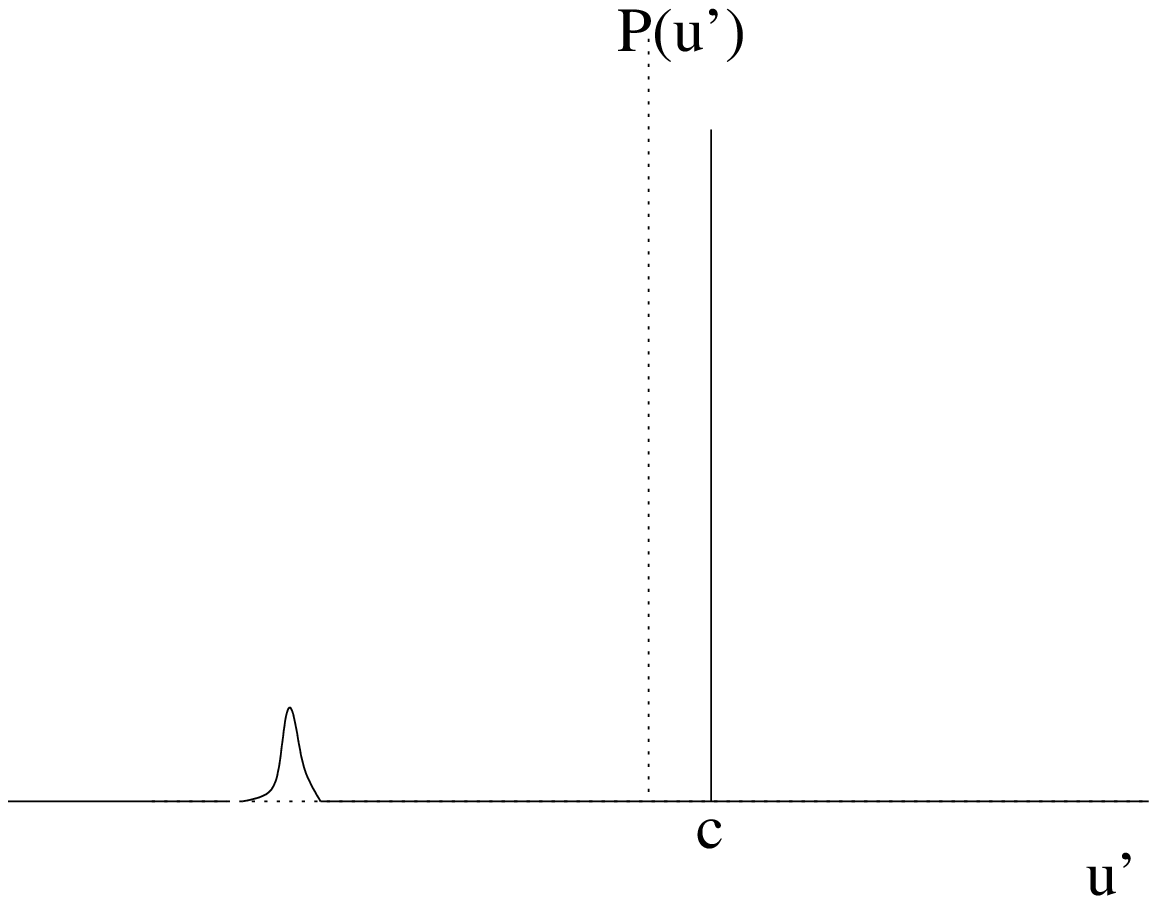,width=12cm,angle=0}}
        \vspace*{0.5cm}
\caption{}
\label{Pnlessfig}
\end{figure}

As mentioned in the introduction, structure function is closely
related to intermittency. Since the exponent $\zeta^{Burg}_q$ is not
proportional to $q$, the noiseless Burgers equation is classified as
intermittent system. However, a point to note is that the velocity of
the noiseless Burgers equation is not random.  From the velocity
profile it is clear that $P(u)= const$ for $u$ between $u_{min}$ and
$u_{max}$.  The slope $u'$ is a constant ($c$) for all $u$ except
within the shock region where $u'$ is large but negative.  Therefore,
$P(u')$ will be a sum of a delta function at $u'=c$ and a small spiky
function at large but negative $u'$ (see Fig.~\ref{Pnlessfig}).  Since
$u$ is not random, it is somewhat confusing to call the signal as
intermittent.  However, it is common practice to classify the
noiseless Burgers equation as an intermittent system.
 
For fluid turbulence it has been shown that the intermittent velocity
field has a multifractal distribution.  This was demonstrated by
Meneveau and Srinivasan \cite{Sree1,Sree2} using the cascade model.
We will apply the same model to Burgers turbulence for a further
understanding of the intermittent nature of the Burgers solution.

In the cascade model of Meneveau and Sreenivasan \cite{Sree1,Sree2}
for fluid turbulence, the nonlinear energy flux $E_r$ is distributed
unequally between two smaller eddies. The flux $E_r$ at length scale
$r$ is divided into fractions $p_1E_r$ and $p_2E_r$ to two smaller
eddies of length $r/2$, and so on. The generalized-dimension of the
resulting multifractal is given by
\begin{equation}
\label{Dq}
D_q=\log _2\{p_1^q+p_2^q\}^{1/(1-q)} .
\end{equation}
The constants $D_q$ are related to the exponents of the structure
function of the energy flux $E_r^q$ in eddy of size $r$ by
\begin{equation}
\sum E_r^q=E_L^q\left( \frac rL\right) ^{\left( q-1\right) D_q} ,
\end{equation}
where the sum is taken over all the eddies at the $n$th stage. When
$p_{1} \neq p_{2}$, the flux function is unevenly distributed after
several bifurcations of the original eddy, and that yields bursty
behavior for $E_r$.  A closer inspection shows that the Burgers
solution corresponds to $p_1\rightarrow 0$ and $p_2\rightarrow 1$,
which implies that $D_q\approx 0$. Hence, for Burgers turbulence, only
one among the $2^n$ eddies has all the flux after $n$ bifurcation of
the original eddy. This corresponds to maximal intermittency or
maximal multiscaling. Using Eq. (5.6) of Meneveau and Sreenivasan
\cite{Sree2}, we get
\begin{equation}
\zeta^{Burg}_q=\left( q/3-1\right) D_{q/3}+1=1,
\end{equation}
a result consistent with Eq. (\ref{zeta_Burg}). Note that when
$p_1=p_2=1/2$, $E_r$ is constant and $D_q=1$, and there is no
intermittency.

In Burgers equation, the dissipation occurs only at the shocks, which
are narrow regions. So, as we traverse along a line, there are regions
of no dissipation, then suddenly a short region (shock) with intense
dissipation appears. According to this observation, the noiseless
Burgers equation exhibits strong intermittency, in fact maximal
intermittency, in the light of Meneveau and Sreenivasan's model
\cite{Sree1,Sree2}.

The structure function of the noiseless KPZ
equation can be easily calculated using the relationship
$h(x,t)=-\int^xu(x^{\prime },t)dx^{\prime }$.  The surface profile
$h(x)$ at a asymptotic time (shown in Fig.~\ref{profilefig}) is
\begin{equation}
h(x,t)=\left\{
\begin{array}{ll}
-\frac{x^2}{2t}+\frac{\eta _i}t\left( x-\xi _i\right) & \mbox{ for
$\frac{1}{2} (\xi_{i-1}+\xi_{i}) < x < \xi_{i}$} \\
-\frac{x^2}{2t}+\frac{\eta _{i+1}}t\left( x-\xi _i\right) & \mbox{ for
$ \xi_{i} < x < \frac{1}{2} (\xi_{i}+\xi_{i+1})$ }
\end{array}
\right.
\end{equation}
After some algebra we can obtain the structure function $T_q(r)$ for
the KPZ equation:
\begin{eqnarray}
\label{zeta_KPZ}
T_q(r) & \approx & \frac{1}{\left( q+1\right) L}\left(
\frac{r}{t}\right)^q \sum_i \left[ \left(\eta _{i+1}-\xi
_{i+1}\right)^{q+1} -\left( \eta _{i+1}-\xi _i\right)^{q+1} \right]
\nonumber \\ & & +\frac{1}{\left( q+1 \right) L}\left(
\frac{r}{t}\right)^q r \sum_i \frac{\left[ \left( \eta _{i+2}-2\xi
_{i+1}\right)^{q+1} -\left(\eta _{i+1}-2\xi _{i+1}\right)
^{q+1}\right] } {\left( \eta _{i+2}-\eta_{i+1}\right) }
\end{eqnarray}
The solution of the KPZ equation has cusps at positions where shocks
appear in Burgers equation.  These cusps yield contributions
proportional to $r^{q+1}$ to the structure function.  Hence, to a
leading order, $T_q(r)\propto r^q$. Therefore, $\zeta^{KPZ}_q=q$.
Note that smooth continuous curves also yield $\zeta_q=q$ because
$h(x+r)-h(x)\approx h' r$ for small $r$.  Since the exponent
$\zeta^{KPZ}_q$ is proportional to $q$, the solution of the noiseless
KPZ equation will be classified to be non-intermittent.  However, note that
the probability density of $h'=-u$ is flat (nongaussion), hence
indicating intermittent behavior for the noiseless KPZ
equation. Clearly there appears to be a contradiction because the
structure function indicates non-intermittency, but the $P(h')$
indicates intermittency.  The resolution of this apparent
contradiction is given below.

The systems under investigation for intermittency have typically a
random output.  In non-intermittent systems, the probability density
of the output is gaussian, i.e., probability of finding a large signal
decreases as $\exp(-x^2)$.  However, in the intermittent systems, the
probability density of the random output signal deviates from the
gaussian behaviour. The key is that the system under investigation for
intermittency must have a random output, which is not the case for the
noiseless KPZ and Burgers equation.  In that sense, it is somewhat
meaningless to ask whether the noiseless Burgers and KPZ equations
exhibit intermittency or not.  This is the reason why the conclusions
from the structure function and those from the probability density
differ for the noiseless KPZ equation.  In literature, however, the
noiseless Burgers equation is classified as an intermittent system.

	In the following section we will derive the energy spectrum of
the noiseless Burgers and KPZ equation using Eqs.~(\ref{struct_Burg},
\ref{zeta_KPZ}).

\section{Energy spectrum of noiseless Burgers and KPZ equation}

We briefly discuss the energy spectrum of the noiseless Burgers and
KPZ equation because they will be compared with the results obtained
for the equations with coloured noise.  We restate the earlier results by
Kida \cite{Kida2}, Gotoh \cite{Goto}, and Bouchaud et al.~\cite{Bouc}.
We can obtain the energy spectrum $E^u(k)=<|u(k)|^{2}/2>$ using
$S_2(r)$.  Clearly,
\begin{equation}
<u(x+r)u(x)>= <u^2> - \frac{1}{2} S_{2}(r).
\end{equation}
Therefore,
\begin{eqnarray}
  	E(k) & = & \frac{1}{L} \int_{-\infty}^{\infty} 1/2
					<u(x+r)u(x)> exp(-i k r) dr
					\nonumber \\ & = &
					\delta_{k,0} \frac{<u^2>}{2} +
					\frac{1}{2 L t^2} \frac{d^2
					\delta_{k,0}}{dk^2} +
					(Lk)^{-2} \sum_{i} \mu_{i}^2.
\end{eqnarray}
  Hence, the noiseless Burgers equation has energy spectrum $E^{u}(k)
\propto k^{-2}$ for $k>0$.  In the above equation, the Fourier
transform of $r^2$ is $d^2 \delta_{k,0}/dk^2 $ \cite{Ligh}.

A similar procedure using $T_2(r)$ will yield $E^h(k)= <|h(k)|^2 /2>$,
\be E^{h}(k)=\delta_{k,0} \frac{<h^2>}{2} + B \frac{d^2}{d k^2}
\delta_{k,0} + \frac{C}{(Lt)^2} k^{-4}, \ee where $B$ and $C$ are
constants \cite{Ligh}.  Hence for $k>0$, we obtain $E^{h}(k) \propto
k^{-4}$ for the noiseless KPZ equation.

Most papers in the past implicitly assume that if $T_{2}(r) \propto
r^{2 \chi}$, then $E^{h}(k)$ should be proportional to $k^{-2 \chi
-1}$.  Clearly this does not hold for the noiseless KPZ equation
(check: $\chi=1$, but $E^{h}(k) \propto k^{-4}$). This apparent
contradictions, which also occurs for $\sigma>1$, can be resolved
using the following arguments.

Suppose $E(k)=A k^{-2 \chi -1}$.  The second-order structure function
$S_{2}(r)$ is given by
\begin{eqnarray}
S_{2}(r) & = & \frac{1}{L} \int_{0}^{L} dx <|u(x+r)-u(x)|^2> \nonumber
	\\ & = & 16 \int_{0}^{\infty} dk E(k) \sin^2 (kr/2) \nonumber
	\\ & = & 16 \int_{0}^{\infty} dk A k^{-2 \chi -1} \sin^2
	(kr/2).
\end{eqnarray}
We are interested in the leading order behaviour of $S_{2}(r)$ for
small $r$.  When $\chi < 1$, the above integral converges, and
\begin{eqnarray}
S_{2}(r) & \propto & 16 A r^{2 \chi} \int_{0}^{\infty} ds s^{-2 \chi
-1} \sin^2 (s) \nonumber \\ & \approx & A C_1 r^{2 \chi},
\end{eqnarray}
where $C_1$ is a dimensionless constant.  When $\chi > 1$, the
integral diverges from below (small $k$).  However, we can cure this
divergence by choosing the lower limit of the integral to be $2
\pi/L$, which yields
\begin{eqnarray}
S_{2}(r) & \propto & 16 A \int_{2 \pi /L}^{\infty} dk k^{-2 \chi -1}
	(kr/2)^2 \nonumber \\ & \approx & A C_2 (r/L)^2 L^{2 \chi},
\end{eqnarray}
where $C_2$ is a constant.

Hence, when $\chi <1$, $E(k) \propto k^{-2 \chi -1}$ and $S_{2}(r)
\propto r^{2 \chi}$ as expected, however when $\chi >1$, $E(k) \propto
k^{-2 \chi -1}$ and $S_{2}(r) \propto r^{2}$ .  This analytical
results are seen in our numerical simulations to be described below.

After the discussion on the noiseless equation, now we turn to the
noisy Burgers and KPZ equations.

\section{Noisy Burgers equation: Various ranges of parameters $\sigma$}

In this section we will attempt to divide the parameter range of
$\sigma$ according to the properties of the solution.  From
Eq.~(\ref{Burgers}) the noise spectrum $|\zeta(k)|^2$ is proportional
to $k^{-2 \sigma}$. Using this we can derive the noise-noise
correlation as
\begin{equation}
\label{Burgernoisex}
\left\langle \zeta (x,t)\zeta (x+r,t)\right\rangle \sim \left\{
\begin{array}{ll}
B_0\delta(r) & \mbox{for $\sigma=0$} \\ 
B_1-C_1 r^{2 \sigma -1} &\mbox{for $0 \le \sigma \le 1/2$} \\ 
B_2-C_2\log(r) & \mbox{for $\sigma=1/2$} \\ 
B_3 - C_3 r^{2\sigma -1} & \mbox{for $1/2 < \sigma \le 3/2$} \\ 
B_4-C_4 r^2 L^{2\sigma -3} & \mbox{for $\sigma >3/2$ }
\end{array}
\right.
\end{equation}
where $L$ is the length of the system, and $B_i$ and $C_i$ are
constants.  For $-1 \le \sigma \le 0$, $|\zeta(k)|^2 \rightarrow
\infty$ for large $k$, hence, the inverse Fourier transform which
yields $<\zeta (x,t)\zeta (x^{\prime},t)>$ is not defined in this
regime. However, Kardar et al. \cite{KPZ} and Medina et
al. \cite{Medi} have solved the KPZ equation for this regime using
$|f(k)|^2$ which is well defined for large $k$.  We call the region
$-1 \le \sigma \le 0$ as (a).

It can be deduced from the above discussion that the noise-noise
correlation increases with the increase of $\sigma$ till
$\sigma=3/2$. Beyond $\sigma=3/2$, the correlation is proportional to
$1-C r^2$ for all $\sigma$. Therefore, it is expected that the $h-h$
and $u-u$ correlation would increase with the increase of $\sigma$
till 3/2, beyond which the behaviour is expected to be somewhat
similar.  Keeping this in mind we have divided the $\sigma$ range
beyond 0 in two regions: (b) $0 \le \sigma \le 3/2$, and (c) $\sigma >
3/2$.  Well defined shocks develop in the parameter range (c) due to
the large noise-noise correlation, and these shocks determine the
exponents. In the intermediate parameter range (b), the exponents
change slowly from RG dominated values to shock dominated values.  The
details are given in the next section.

Polyakov \cite{Poly} has analytically solved Burgers equation with
$\sigma=3/2$ using the methods of quantum field theory.  Boldyrev
\cite{Bold1,Bold2} extended Polyakov's method to the range $1/2 \le
\sigma \le 3/2$ and derived the spectral exponents and the probability
densities.  Boldyrev derived that $\chi = 1+2 \sigma /3$ for $1/2 \le
\sigma \le 3/4$, and $\chi=3/2$ for $\sigma > 3/4$.  Boldyrev
\cite{Bold1} suggests that the above formula may also be valid for $0
\le \sigma \le 1/2$.  In the following section we will compare our
numerical results with the theoretical predictions of Medina et
al. \cite{Medi}, Polyakov \cite{Poly}, and Boldyrev
\cite{Bold1,Bold2}.

\section{Simulation method}
 
Our calculations in this paper have been done using direct numerical
simulations based on pseudo-spectral method. This method, commonly
used in turbulence simulations, is expected to perform better than
finite difference scheme because the derivatives can be calculated
exactly in the spectral method \cite{Canu}.  The finite difference
scheme was adopted by Moser et al. \cite{Mose} and Peng et al. \cite
{Peng} in their simulation of KPZ equation for $-1 \le \sigma \le 0$.

We solve the KPZ equation in one dimension. The details of the
simulation are as follows.  A box of size $2\pi $ is discretized into
$N=1024$ divisions.  The KPZ equation is solved in Fourier
space. However, to compute the nonlinear term, we go to real space,
perform multiplication, then again come back to Fourier space.  We
time advance the Fourier components $h(k,t)$ using Adam Bashforth time
marching procedure with flat surface as an initial
condition. Two-third rule is used to remove aliasing \cite{Canu}. For
details of the simulation refer to Canuto et al.  \cite{Canu} and
Verma et al. \cite{MKVmhdsim}. In our simulation we also add
hyperviscosity term $(\kappa \nabla ^4h)$ to the KPZ equation to damp
the large wavenumber modes strongly. The hyperviscosity term does not
affect the intermediate scales which is of our interest
\cite{MKVmhdsim}; this is because the $\nabla ^4h$ and higher order
derivative terms are irrelevant in the renormalization group sense
(cf. Barabasi and Stanley \cite{Bara} and references therein).

In our simulation we take $\nu$ and $\kappa$ to be very small.  Note
that large $\nu$ correspond to Edward-Wilkinson (EW) equation. The
dimensionless parameters used in our simulations are
\begin{eqnarray}
\label{para}
  \lambda & = & 1.0, \nonumber \\ \nu & = & 10^{-5}, \nonumber \\
 \kappa & = & 10^{-6}, \nonumber \\ (2 \pi)^{2}D & = & 10^{-3},
 \nonumber \\ dt & = & 1/2000
\end{eqnarray}
with one exception.  For $\sigma=-1$, we choose $\lambda=0.1$ for the
stability of the code.  In the Appendix A we have varied values of the
parameters $\nu$ and $D$ and shown crossover from KPZ behaviour to EW
equation.

The values of $\sigma $ used in our simulation are -1, -0.85, -0.75,
-0.60, -0.50, -0.25, 0, 1.25, 1.50, 1.75, 1, 2, and 6.  We have time
evolved the equation till 15 nondimensional time units.  We find that
the system reaches saturation in approximately 8 to 10 time units.
For reference, $2 \pi$ time unit corresponds to one eddy turnover time
in fluid turbulence. For ensemble averaging, we have performed
averages over 100 samples which start with different random seeds for
the noise. We have used $ran1$ of numerical recipes \cite{Recipe} as
our random number generator.  Each computer run for 15 time units and
100 samples takes approximately 7 hours on a Pentium machine (150
MHz).

\begin{figure}
\centerline{
        \psfig{figure=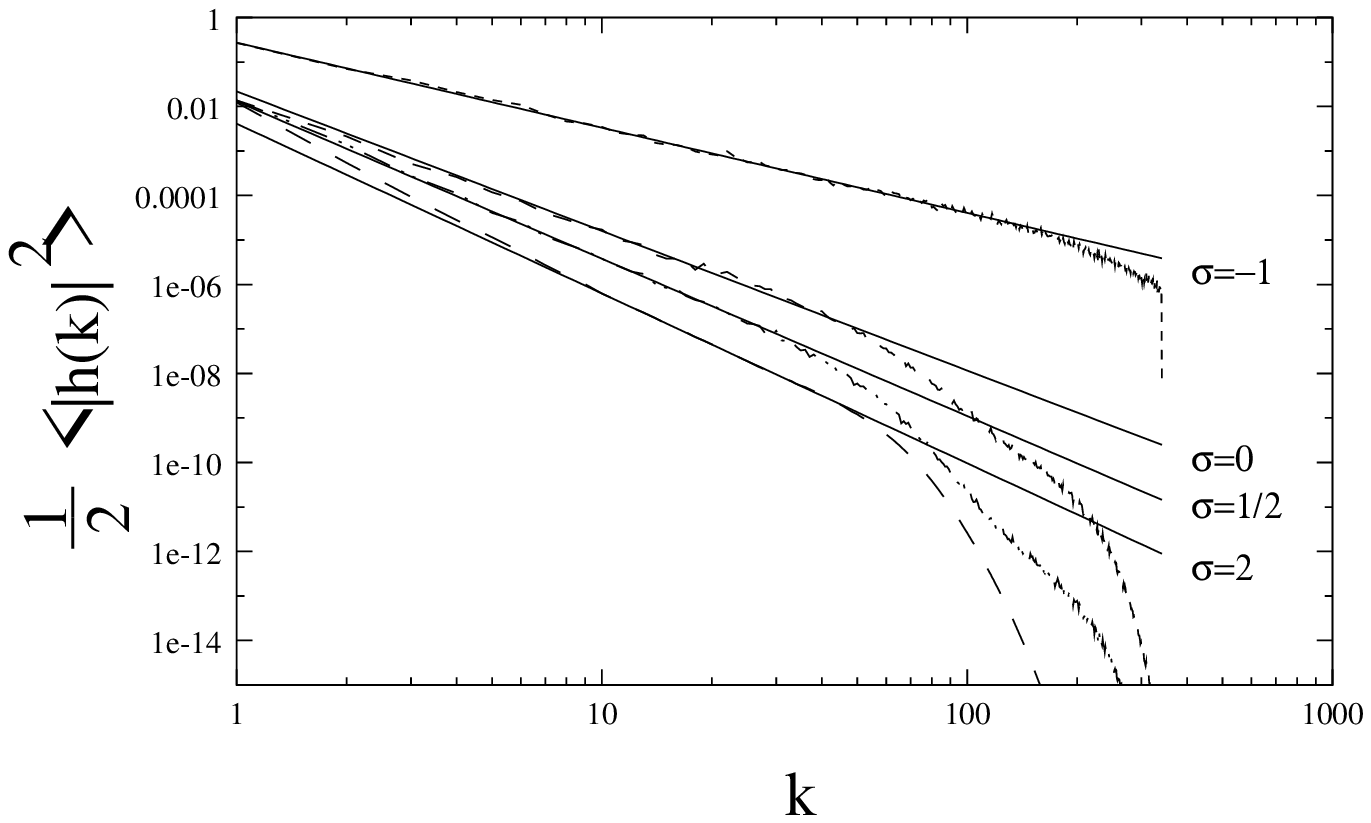,width=12cm,angle=0}}
        \vspace*{0.5cm}
\caption{}
\label{ekfig}
\end{figure}

We have calculated $\chi $ and $\beta $ using the simulation data. The
ensemble average $<.>$ have been obtained by taking averages over 100
runs.  The width $W(L,t)=[<\sum_x h^2 (x)/N>]^{1/2}$ grows as a 
power law in time,
i.e., $W(L.t) \propto t^{\beta_{KPZ}}$, in the early stages of growth.
We obtain $\beta_{KPZ}$ by fitting a straight line in $log-log$ plot
of $W(L,t)$ vs. $t$ over a range of $t=0.2:2.5$.  The other exponent
$\chi_{KPZ}$ is obtained from the asymptotic $<|h(k,t)|^2>$ averaged
over $100$ runs. We find that
powerlaw holds  ($|h(k)|^2\propto k^{-2\chi_{KPZ} -1}$)  for $k=10:8$; 
this range corresponds to the inertial
range of turbulence. We perform the averaging at steady-state ($t=15$).
In Fig.~\ref{ekfig} we plot $<|h(k)|^2>$ vs. $k$ for $\sigma =-1,0,1/2,$
and $2$. The lines of best fit are also shown in  Figure.  The
computed values of the exponents are listed in Table 1. Our estimate
of the error in the exponent is roughly 0.05.

We have also calculated the structure functions $S_q(r)$ and $T_q(r)$
using the simulation data. The variable $r$ ranges from 1 to $N/2$.
Near the boundaries we calculate $\Delta u$ or $\Delta h$ using wrap
around scheme, i.e., $x+r$ is taken as $mod(x+r,N)$.  we have reported
the structure function exponents at the steady-state ($t=15$).

The probability densities $P(h), P(u)$, and $P(u')$ have been
calculated by averaging the histograms over 100 runs at every 0.2 time
interval from initial time of 5 units to the final time of 15 units.
Even though the solution has not reached the steady state at $t=5$,
for more sampling we have taken the time interval from $t=5$ to
$t=15$.

In the following section we will describe the results of our
simulation for various degrees of noise correlations, i.e., for
different $\sigma$s.

\section{Simulation results}

As discussed in section IV, we divide the range of parameter $\sigma$
in three regions: (a) $-1 \le \sigma \le 0$, (b) $0 \le \sigma \le
3/2$, and (c) $\sigma \ge 3/2$.  Fig.~\ref{uxpapfig} shows numerical
$u(x,t)$ for various $\sigma$s.  We find that noise dominates when
$\sigma$ is negative.  However, shock-like structures are clearly
visible for $\sigma=2,6$ (compare with Fig.~\ref{profilefig}).  In the
the intermediate range of $\sigma$ (region (b)), both noise and shock
structure coexist.  We show below that the structures present in the
profile contribute significantly to the determination of 
the spectral and intermittency exponents.

Regarding energy spectrum, our numerical results are in good agreement
with the RG predictions for $-1 \le \sigma \le 0$ or $0 \le \rho \le
1$.  There is a gradual transition from $\chi_{KPZ}=1$ to
$\chi_{KPZ}=3/2$ as $\sigma$ increases from 0 to 3/2. For $\sigma \ge
3/2$ we find that $\chi_{KPZ}=3/2$.

Regarding probability density, we find that the probability densities
$P(h)$ and $P(u)$ are gaussian for all $\sigma$'s.  Fig.~\ref{Pufig}
shows the plot of $P(u)$ vs. $u$ along with the best fit gaussian
curves.  Clearly $P(u)$ is gaussian for all the $\sigma$s.  Therefore,
the exponents of the structure functions $T_q(r)$ are expected to be
proportional to $q$.

Fig.~\ref{structfig} shows a typical log-log plots of $S_q(r)$ and
$T_q(r)$ vs. $r$ (here shown for $\sigma=6$).  The shock region is
$r<10$ where both $S_q(r)$ and $T_q(r)$ are proportional to $r^q$
because $u(x)$ and $h(x)$ are continuous here.  The region of our
interest is $r=10:80$ where both $S_q(r)$ and $T_q(r)$ are power laws.
The exponents for this range of $r$, $\zeta^{Burg}_q$ and
$\zeta^{KPZ}_q$, are listed in Tables 1 and 2 respectively.
Fig.~\ref{kpzfig} shows $\zeta^{KPZ}_q$ vs. $q$ plot, and
Fig.~\ref{burgfig} shows $\zeta^{Burg}_q$ vs. $q$ plots.  Clearly,
$\zeta^{KPZ}_q$ is approximately proportional to $q$ 
as predicted in the previous paragraph.
Hence, our results regarding $P(u)$ and $T_q(r)$ are consistent.

We find, however, that $P(u')$ deviates significantly from the
gaussian behaviour as we increase $\sigma$ from 0, thus signalling an
intermittent behaviour for $u$.  The details of our results for
various ranges are given below.

\subsection{$-1 \le \sigma \le 0$}

In Table 1 we have listed our numerical values of $\chi_{KPZ}$ and
$\beta_{KPZ}$.  For comparison we also list the predicted values of
Medina et al. \cite{Medi} below
\begin{equation}
\chi =\left\{
\begin{array}{ll}
\frac 12 & \mbox{for $-1 \leq \sigma \leq -\frac{3}{4}$ } \\ 1+\frac
23 \sigma & \mbox{for $-\frac{3}{4} \leq \sigma \leq 0$ }
\end{array}
\right.
\end{equation}
and
\begin{equation}
\beta =\left\{
\begin{array}{ll}
\frac 13 & \mbox{for $-1 \leq \sigma \leq -\frac{3}{4}$} \\
\frac{3+2\sigma }{3-2\sigma } & \mbox{for $-\frac{3}{4} \leq \sigma
\leq 0$ }
\end{array}
\right.
\end{equation}
Medina et al.'s results are based on RG scheme that breaks down beyond
$\sigma=0$.

We compute $\chi_{KPZ}$ by a straight line fit to the $\log-\log$ plot
of $<|h(k)|^2>$ over $k=10:80$ at $t=15$ (see Fig.~\ref{ekfig}).  Note
that $|h(k)|^2\propto k^{-2\chi-1}$.  The $\chi_{KPZ}$s listed in
Table 1 show that that our numerical $\chi_{KPZ}$s are in close
agreement with Medina et al.'s \cite{Medi} and Chattopadhyay and
Bhattacharjee's \cite{ChatJKB} theoretical predictions.  Our findings
are also consistent with earlier simulation results by Amar et
al. \cite{Amar}, Peng et al. \cite{Peng}, and Meakin and Jullien
\cite{Meak1,Meak2}.  For $\sigma=-1/4$, Hayot and Jayaprakash
\cite{Hayo} have reported a crossover in the wave number space---from
RG dominated region for small $k$ to free field behaviour for large
$k$.  We do not find any such crossover in our simulation.  Regarding
the Burgers equations, $\chi_{Burg}=\chi_{KPZ}-1$.

Regarding $\beta_{KPZ}$ calculations, our results are agreement with
Medina et al.'s predictions for $\sigma=-1:-3/4$.  However, our
exponents differ significantly with Medina et al.'s exponents for
higher $\sigma$s.  For example, for $\sigma=0$, we obtain
$\beta_{KPZ}=0.45$ contrary to the predicted $\beta_{KPZ}=1$.  The
reason for this discrepancy is not clear to us at this
point. Regarding $\beta_{Burg}$, we find that it is approximately 0.5,
consistent with Hayot and Jayaprakash's findings \cite{Hayo}.

The probability density $P(u')$ vs. $u'$ for $\sigma=-1,-1/2,0$ is
shown in Fig.~\ref{P0to1fig} on a semilog plot.  The figure shows that
$P(u')$ is gaussian for $\sigma=-1$, but it deviates from the gaussian
behaviour with the increase of $\sigma$; the deviations becomes more
and more prominent for higher $\sigma$.  At $\sigma=0$, $P(u') \approx
0.01*\exp(-0.14 |u'|)$ for negative $u'$ (the curve of the best fit
shown in the Fig.~\ref{P0to1fig}).  Beyond $\sigma=0$, a powerlaw
tails start appearing for negative $u'$.

Table 2 and Fig.~\ref{burgfig} show that $\zeta^{Burg}_2$ is close to
zero for $\sigma < 0$.  The energy spectrum $ <|u(k)|^2/2> \propto
k^{- 2 \chi_{KPZ}+1}$.  Therefore, using the arguments of section 3 we
can easily show that $<| u(x+r)-u(x)| ^2> \propto r^{2(\chi_{KPZ}-1)}$
for $\sigma < 0$.  For $\sigma=0$, it can be easily shown that $<|
u(x+r)-u(x)| ^2> \propto \log (r)$.  We find in our simulation that
$\zeta_2^{KPZ}$ is positive contrary to the above prediction.
This deviation may be because of {\em intermittency} (due to the
presence of small shocks).  It is also possible that at large $r$ ($r
\approx L$), the other sub-critical terms may become comparable to the
leading order term and may change the exponent.  Quantitative
calculation of the exponents for this range of $\sigma$ is beyond the
scope of this paper.

\subsection{$\sigma \ge 3/2$}

\begin{figure}
\centerline{
        \psfig{figure=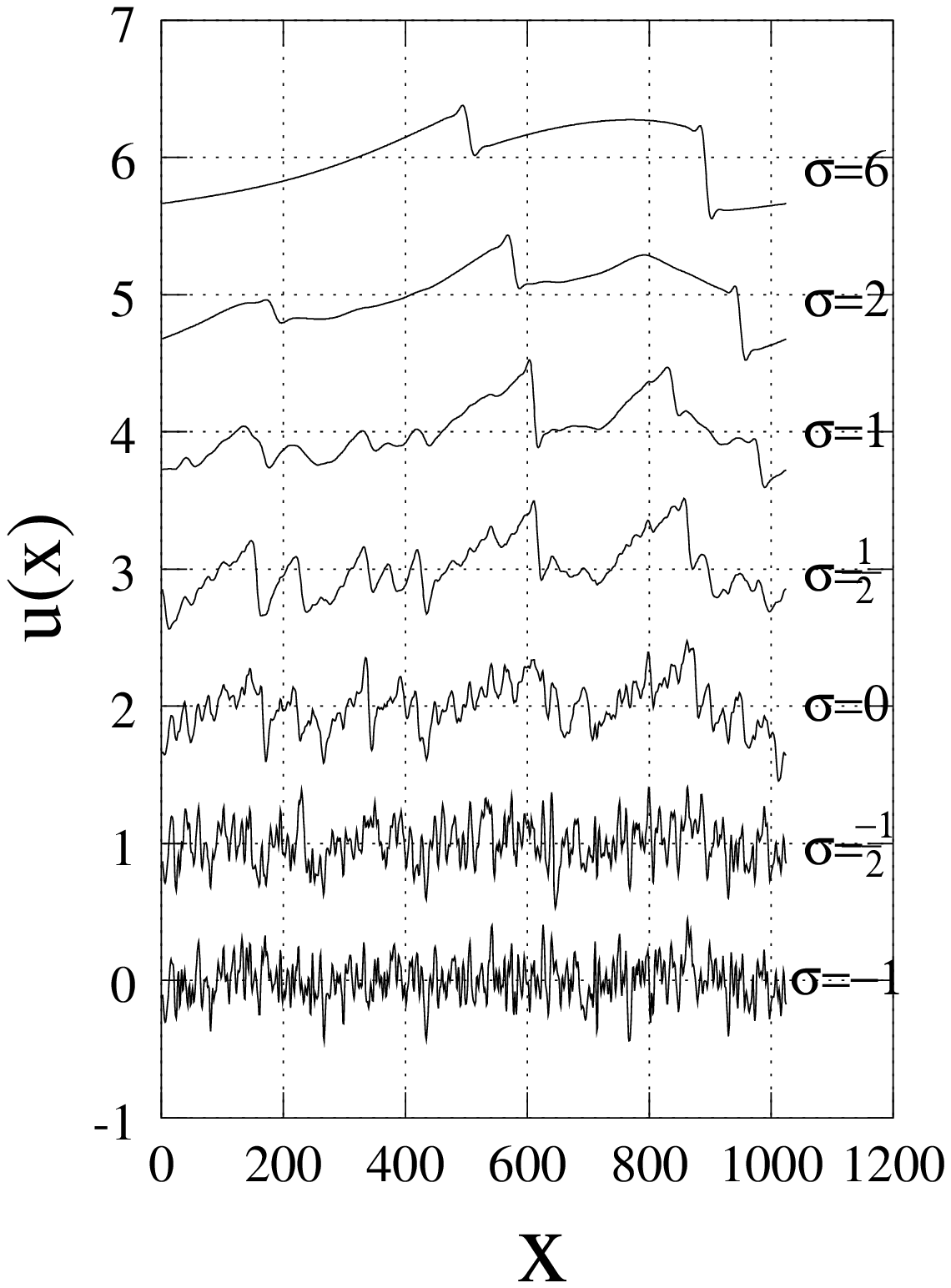,width=12cm,angle=0}}
        \vspace*{0.5 cm}
\caption{}
\label{uxpapfig}
\end{figure}

\begin{figure}
\centerline{
        \psfig{figure=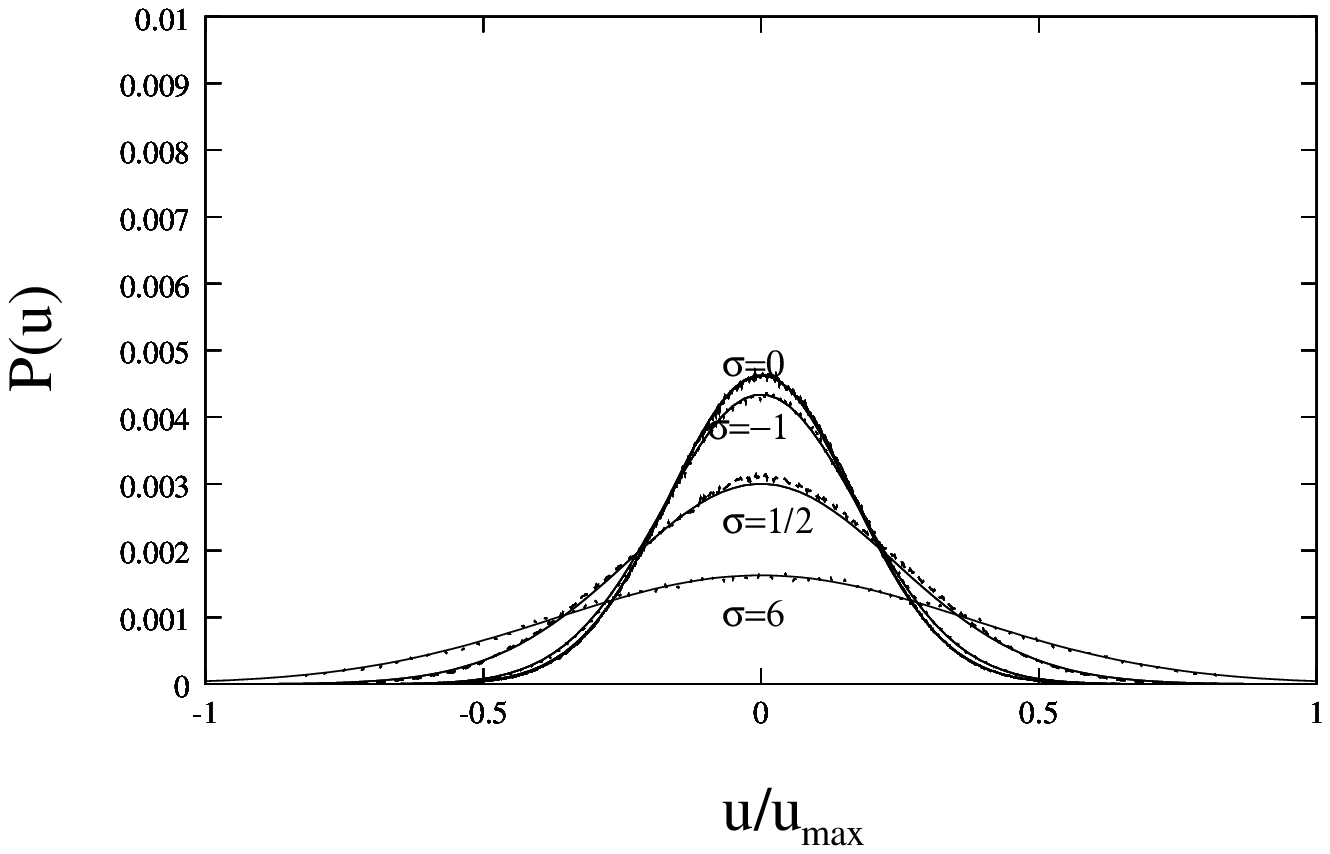,width=12cm,angle=0}}
        \vspace*{0.5 cm}
\caption{}
\label{Pufig}
\end{figure}

\begin{figure}
\centerline{
        \psfig{figure=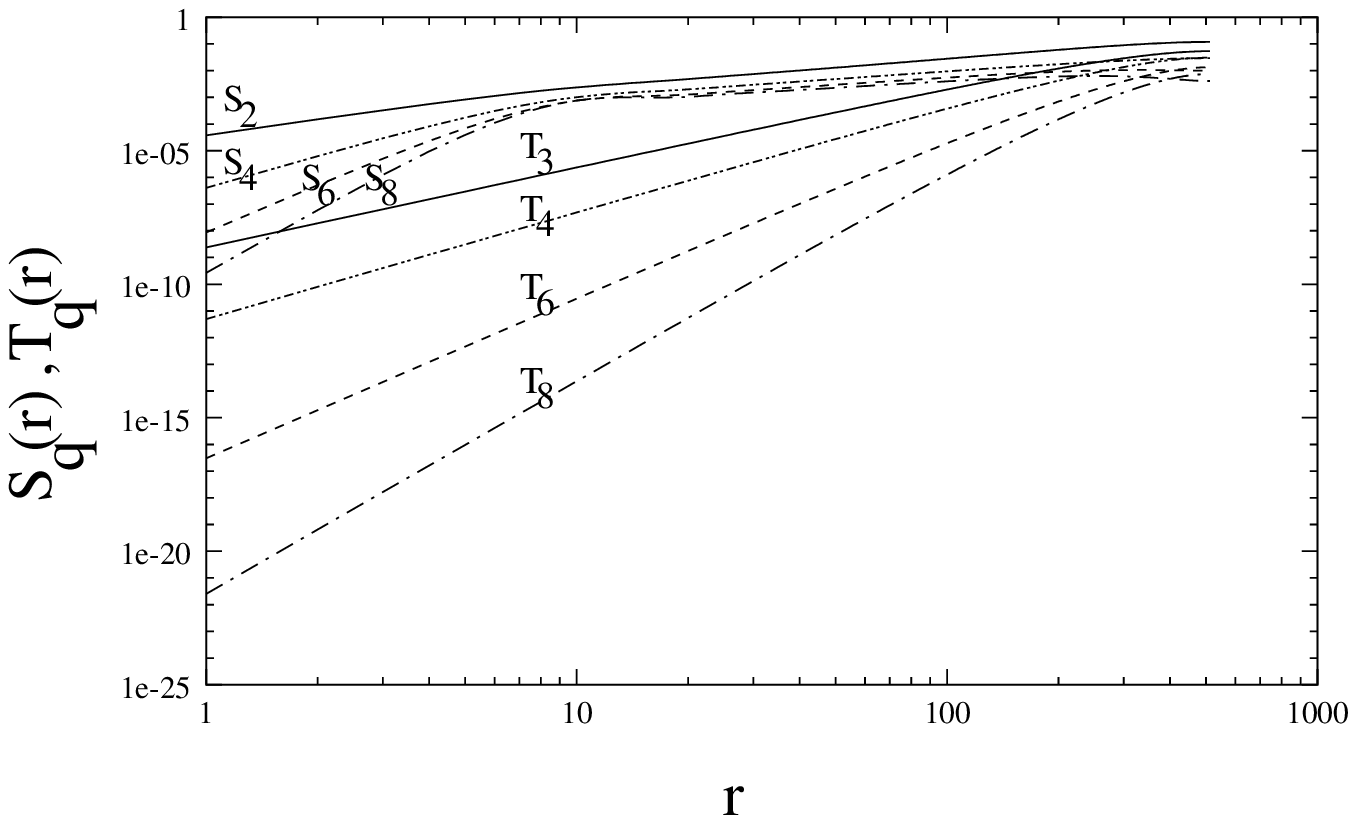,width=12cm,angle=0}}
        \vspace*{0.5 cm}
\caption{}
\label{structfig}
\end{figure}

\begin{figure}
\centerline{
        \psfig{figure=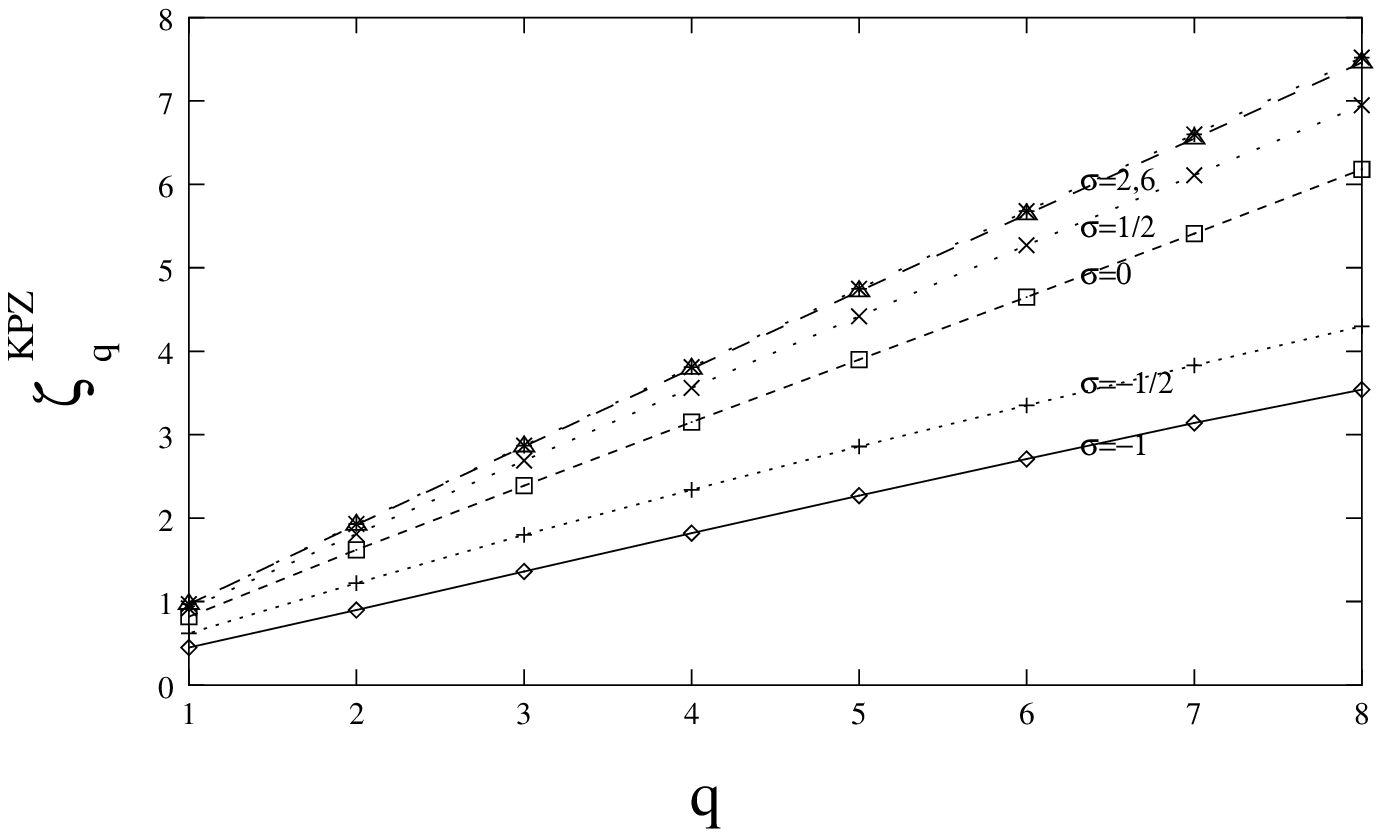,width=12cm,angle=0}}
        \vspace*{0.5 cm}
\caption{}
\label{kpzfig}
\end{figure}

\begin{figure}
\centerline{
        \psfig{figure=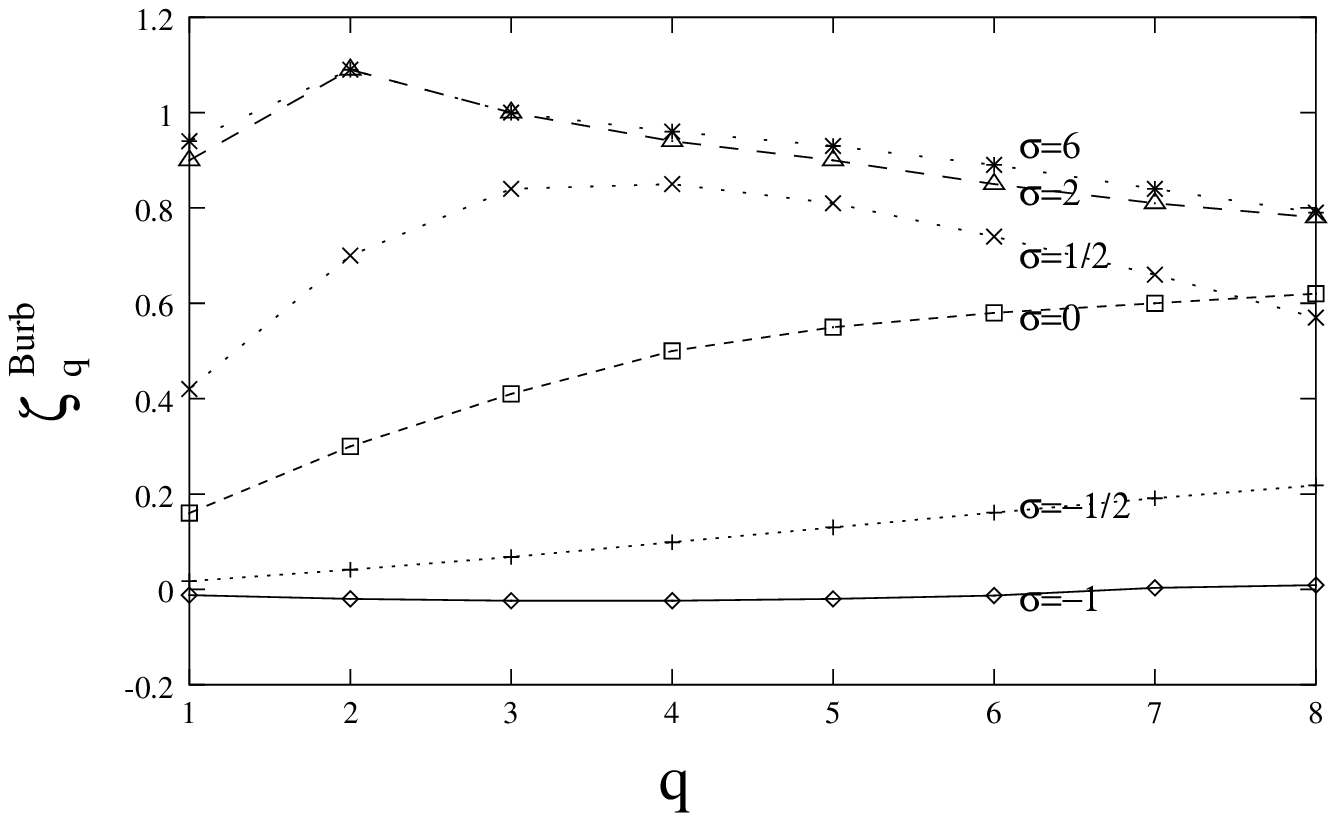,width=12cm,angle=0}}
\vspace*{0.5 cm}
\caption{}
\label{burgfig}
\end{figure}

\begin{figure}
\centerline{
        \psfig{figure=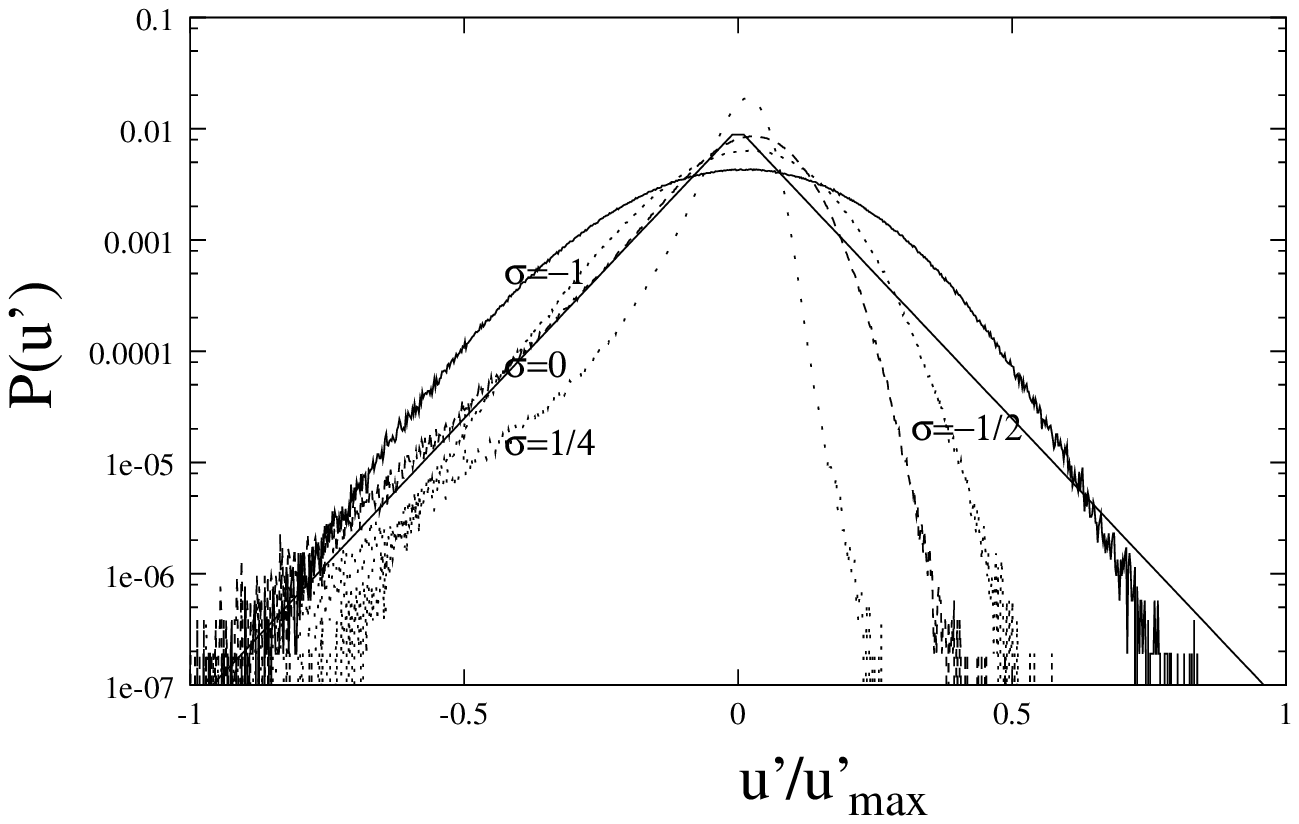,width=12cm,angle=0}}
        \vspace*{0.5 cm}
\caption{}
\label{P0to1fig}
\end{figure}

One of the important aspect of this paper is the discussion of the
energy spectrum, probability density and structure functions of the
Burgers equation with large $\sigma$. As shown in the
Fig.~\ref{uxpapfig}, shocks are prominent for this range of $\sigma$.
Due to these shocks, the exponent $\chi_{KPZ} \approx 3/2$ for $\sigma
> 3/2$ (see Table 1 and Fig.~\ref{ekfig}).  Note that $\chi_{KPZ}=3/2$
and $\chi_{Burg}=1/2$ for the noiseless Burgers equation due to the
presence of shocks.

The existence of shocks for large $\sigma$ can be argued from the
noise-noise correlation discussed in section IV.  For large $\sigma$,
there is long-ranged noise-noise correlation. Physically, large fluid
parcels are moved around by this noise.  As a consequence, there will
be regions where the parcels forced by appositely directed noise
collide with each other and create ``strong'' shocks.  Hence, it is
not surprising that strong shocks will be generated for large $\sigma$.
These shocks determine the dynamics and the spectral indices of
Burgers and KPZ equations.  Therefore, the noisy KPZ equation with
large $\sigma$ yields the same energy spectrum and multiscaling
exponents as the noiseless KPZ equation.

The noise-noise correlation is proportional to $1-C r^2$ for all
$\sigma \ge 3/2$, hence the spectral and multiscaling exponents are
expected to be somewhat similar for all the $\sigma$ beyond 3/2.  This
is borne out by our numerical simulation.  This is consistent with the
trivial observation that for large $\sigma$ only $k=1$ mode is
effective.  It is interesting to note that $k=1$ mode (i.e.,
$\zeta(x)=sin(x)$) also yields noise-noise correlation proportional to
$1-C r^2$, and approximately the same exponents as those 
with $\sigma \ge 3/2$.

We have also analyzed the stability of the shocks at a preliminary
level.  For large $\sigma$, the significant contribution to the
dynamics comes only from the $k=1$ mode of the noise.  Since the first
mode ($k=1$) is noisy, the the formation of a single shock delayed for
some time because of the movement of the zero of the noise
signal. However, after the shock is formed, it shifts around by only a
small amount because the impulse due to the random noise is not strong
enough to move the shock by a large distance.

We have also calculated $\beta$ for both Burgers and KPZ equations
when $\sigma>3/2$ and found them to be approximately 1/2 .  This result
is consistent with findings of Hayot and Jayaprakash \cite{Hayo}.

Regarding the intermittency exponents, as shown in Fig.~\ref{burgfig},
$\zeta_q^{Burg} \approx 1$ for all $q$ for $\sigma
\ge 3/2$.  This result is consistent with existence of shocks (see
Fig.~\ref{uxpapfig} and section II).     For large $r$, $S_q(r)$ is not 
proportional to $r$.  This is  because of the term similar to
$(\mu_{i}-r/t)^{q}$  of Eq.~(\ref{struct_Burg}), which becomes
important at large $r$.  The effects of this subcritical
term is evident in the intermittency exponents listed in
Table 2.  We had anticipated $\zeta_q^{Burg}$ to be 1, but we
consistently find them to be less that 1.  The same trends were
observed for the noiseless case.

\begin{figure}
\centerline{
        \psfig{figure=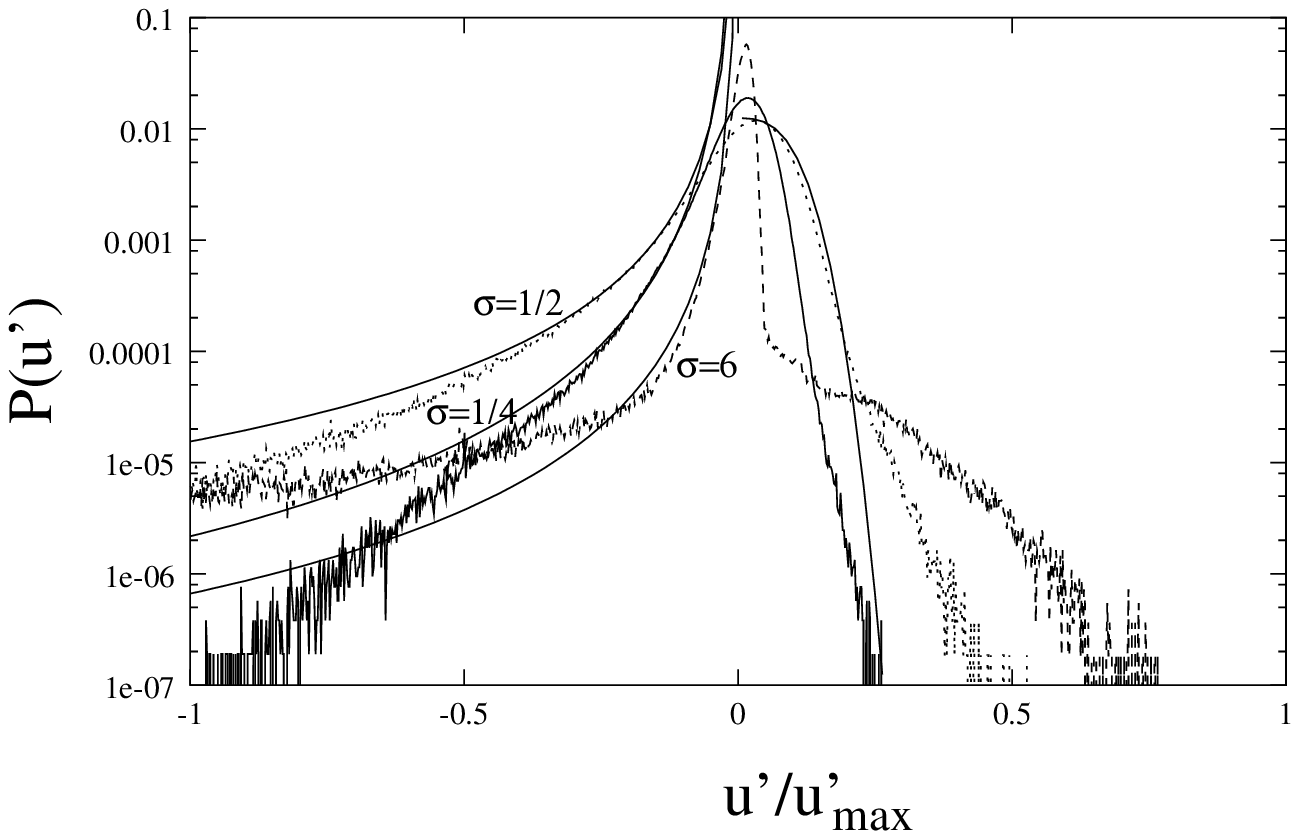,width=12cm,angle=0}}
        \vspace*{0.5 cm}
\caption{}
\label{P1gtfig}
\end{figure}

Fig.~\ref{P1gtfig} shows the probability density $P(u')$ for various
$\sigma$s. The curves of best fit are also shown in the Figure.  The
probability density is clearly nongaussian for all $\sigma$s shown in
the Figure.  In our simulations we find that for all $\sigma \ge 3/2$,
$P(u') \propto u'^{-\alpha} (\alpha=2.2-2.5)$ fits quite well with the
numerical $P(u')$ for the intermediate range of the negative $u'$.
This result is in agreement with the Polyakov's theoretical prediction
for $\sigma \ge 3/2$ \cite{Poly}.  Curiously, $P(u')$ for $u'>0$ is
also nongaussian, but it does not match with the Polyakov's
predictions for positive $u'$.

We have shown in this section that both noiseless Burgers equation and the
Burgers equation with strongly correlated noise have $S_q(r) \propto
r$.  However, the probability densities $P(u)$ and $P(u')$ are very
different for these two cases (see Figs.~\ref{Pnlessfig} and
\ref{P1gtfig}).  This difference is because of the different nature of
the noise.

The entries in Table 2 show that $\zeta^{KPZ}_q=q$.  This can be
argued as follows. Since $\chi_{KPZ} \approx 3/2$, the second order
structure function $T_2(r)$ will be proportional to $r^2$, or
$\zeta^{KPZ}_2=2$ (see Section III). This is seen in our simulation for
$\sigma \ge 3/2$.  The Fig.~\ref{Pufig} also shows that $P(u)$ is
gaussian.  Therefore, $T_q(r) \propto [T_2(r)]^{q/2} \propto r^q$,
hence $\zeta^{KPZ}_q=q$, a result seen in our numerics.  Thus,
structure function and energy spectrum calculations are consistent.

After discussing the shock dominated region of $\sigma \ge 3/2$, we
turn to the region $0 \le \sigma \le 3/2$.

\subsection{$0 < \sigma \le 3/2$}

The $\chi_{KPZ}$ calculated by our numerical simulation for the range
$0 \le \sigma \le 3/2$ is listed in Table 1.  The exponent increases
from 1 and saturate at 3/2.  That is, there is a gradual shift from RG
dominated exponents to shock dominated exponents as we vary $\sigma$.
The profile $u(x,t)$ of Fig. 4 shows that both fluctuations and
shocks coexist in this range, with shocks becoming more and more
important as $\sigma$ increases.

Boldyrev \cite{Bold1,Bold2} has calculated $\chi_{KPZ}$ using methods
of quantum field theory and derived that $\chi_{KPZ} = 1+2 \sigma /3$
for $1/2 \le \sigma \le 3/4$, and $\chi_{KPZ}=3/2$ for $\sigma > 3/4$.
Boldyrev \cite{Bold1} suggests that the above formula may be valid for
$0 \le \sigma \le 3/2$.  Comparison of Boldyrev's predictions with the
entries of Table 1 shows that the predictions work quite well till $0
\le \sigma \le 1/2$.  After that there is a significant deviation, and
the exponents appear to be outside numerical error bars. We need to
probe the region $1/2 \le \sigma \le 3/4$ carefully to reach a
definite conclusion.

We presume that both fluctuations and the embedded structures are
important in the determination of the spectral index.  One would need
to combine contributions from the structures and fluctuations to
obtain the energy spectrum \cite{Supr1}.  In Appendix B we sketch an
elementary framework when both structure and fluctuations are present
in the system.

The Burgers equation for $\sigma=1/2$ has been studied extensively by
Chekhlov and Yakhot \cite{Chek1}.  They find in their high resolution
simulation that $|u(k)|^{2} \propto k^{-5/3}$.  The spectrum in our
low resolution simulation is $|u(k)|^{2} \propto k^{-1.54}$
($\chi_{KPZ}=1.27$), which is close to Chekhlov and Yakhot's
\cite{Chek1} result.  Chekhlov and Yakhot \cite{Chek1} have argued for
a constant cascade of energy in the wavenumber space in this case and
claimed that the Kolmogorov-like energy spectra is due to the
constancy of the energy flux.  However, we find the cascade rate to be
constant for all $\sigma> 1/2$, but the spectral index of the energy
spectrum is in the range of 5/3 to 2 (to be discussed in Section 7).
Hence, the argument that the constant flux yields Kolmogorov's
spectrum is incorrect.

Regarding the probability distribution, in Fig.~\ref{P1gtfig} we plot
 $P(u')$ vs. $u'$ along with the curves of best fit.  It is found that
 $P(u')$ is a powerlaw for $u'<0$. For $\sigma=1.25, P(u') \propto
 u'^{-2.86}$, but for $\sigma = 1/2$, $P(u') \approx u'^{-2.2}$, with
 an error of approximately $10\%$ for both the cases.  Curiously,
 Polyakov's predictions ($\sigma=3/2$) for positive $u'$ appears
to be applicable only for $\sigma=1/2$;  
here $P(u') \propto \exp(-C x^3)$.

The roughening exponents
$\chi_{KPZ}$ lies between 1 and 3/2. Therefore,
$\zeta^{KPZ}_q$ should be proportional to $q$ as argued in the
previous subsection.  Our numerical simulations yield approximately
the same exponents (see Table 3).  Hence our energy spectrum
calculations and the structure function calculations 
are consistent.

\subsection{Summary of simulation results}

From the above discussion we see that Burgers and KPZ equations are
well understood for $-1 \le \sigma \le 0$ and $\sigma > 3/2$.  In the
intermediate range $0 < \sigma < 3/2$, Boldyrev's predictions
appear to explain the numerical data in part of the regime;
further analytic and numerical work is required in this regime.

For all $\sigma > -1/2$, $\chi_{KPZ}+z$ $(z=\chi_{KPZ}/\beta_{KPZ})$
deviates significantly from 2.  In fact, for larger $\sigma$,
$\chi_{KPZ}+z=3/2+(3/2)/(1/2)=9/2$, quite far from 2.  Note that
Medina et al. \cite{Medi} argue that  the identity $\chi_{KPZ}+z=2$ is 
a consequence of Galilean invariance.  However, Hayot and Jayprakash
\cite{Hayo}, Polyakov \cite{Poly} and others have speculated violation
of this identity due to the presence and motion of the shocks.  Meakin and
Jullien's \cite{Meak1,Meak2} results also violate $\chi_{KPZ} +z=2$
condition for a range of $\sigma $. They find that for $\sigma =-1/4,$
$\chi_{KPZ}+z=2.37$, quite different than 2.
  The violation of Galilean invariance even in
the region $-1 \le \sigma \le 0$ signals importance of structures
\cite{Poly}.

We also show that $P(u)$ is gaussian for all
$\sigma$.  However, $P(u')$ is gaussian at $\sigma=-1$, but continuously
changes to $\exp(-x)$ then to a powerlaw behaviour.  
Therefore, $u$ field of Burgers equation exhibits intermittency.
The degree of intermittency depends on $\sigma$.

The energy flux play an in important role in turbulence analysis. In the
following section we briefly 
report energy flux studies for the Burgers equation.

\section{Energy flux of noisy Burgers equation}

We derive an equation which gives us the energy transfer from the
region $|k| \le K$.  The derivation of the energy equation from 
Eq. (\ref{Burgers}) and averaging yields \cite{McCobook}
\be 
\label{flux_eq}
\frac{\partial}{\partial t} \int_{0}^{K} E(k)dk = 
	- \int_{0}^{K} 2 \nu k^2 E(k) dk \\
	- \int_{0}^{K} \Re <u^*(k) \left[ FT
	\left(\frac{\partial}{\partial x} u^2 /2 \right) \right]_k>
	+ \int_0^K \Re <u^*(k) f(k)>.
\ee
From this equation, clearly the energy dissipation in the wavenumber 
sphere of radius $K$ is 
\be
D_{K}= \int_{0}^{K} 2 \nu k^2 E(k) dk, 
\ee
and the energy flux coming out the wavenumber sphere of radius $K$ is 
\be
\Pi_{k}=\int_{0}^{K} \Re <u^*(k) \left[ FT
 \left(\frac{\partial}{\partial x} u^2 /2 \right) \right]_k >.
\ee
It can easily seen that for the random noise, the energy supplied by
the forcing to the wavenumber sphere of radius $K$ is \cite{McCobook}
\begin{eqnarray}
\label{FK}
F_{K} & = &  \int_0^K \Re <u^*(k) f(k)> \nonumber \\
	& = & \int_0^K <|f(k)|^2>. 
\end{eqnarray}
From the Eq. (\ref{flux_eq}) it is clear that at the steady state
\be
F_{K}=\Pi_{K}+D_{K}.
\ee

\begin{figure}
\centerline{
        \psfig{figure=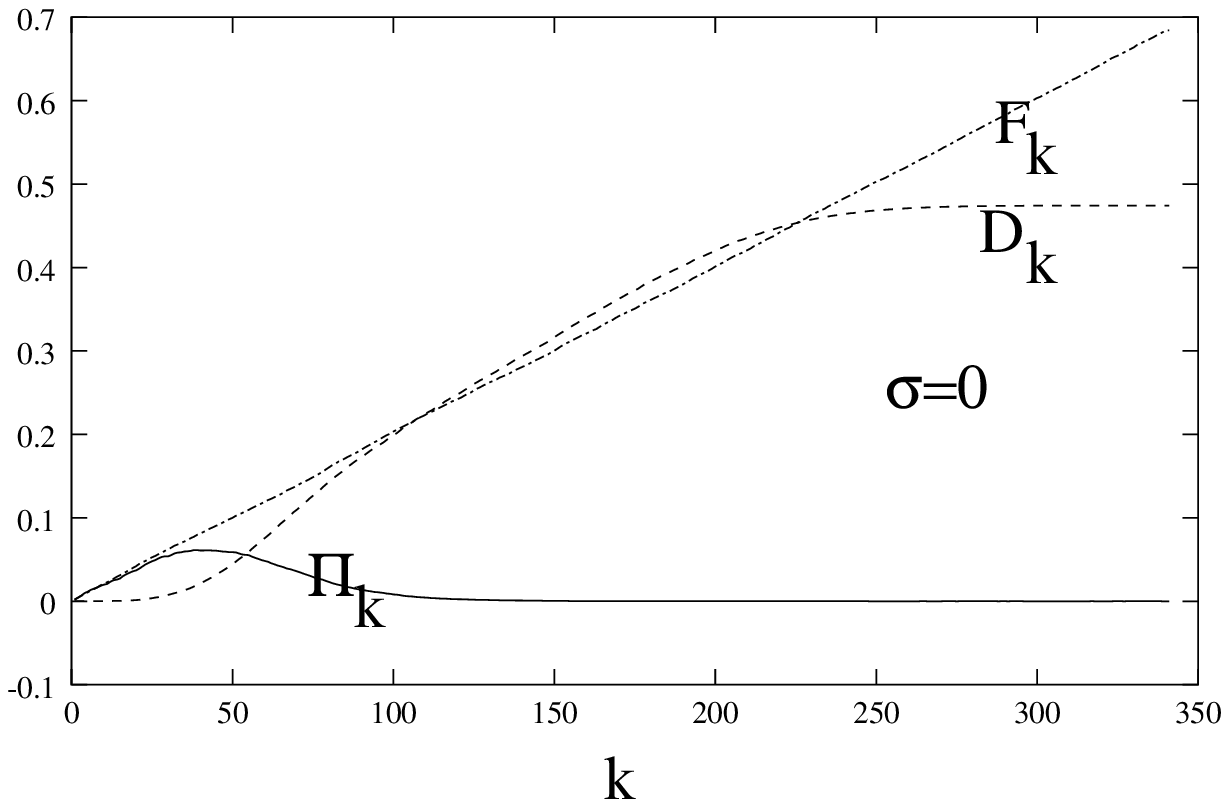,width=12cm,angle=0}}
        \vspace*{0.5cm}
\caption{}
\label{flux0fig}
\end{figure}

We have plotted $D_{K}, \Pi_{K}$, and $F_{K}$ for various $\sigma$.
Fig.~\ref{flux0fig} shows the plots for $\sigma = 0$.  Here, we find 
that $D_{K}
\approx F_{K}$ till $k \approx 250$, but $\Pi_{K} \ll F_{K}$ except
for small $k$.  Similar results are obtained for all $-1 \le \sigma \le
0$.

\begin{figure}
\centerline{
        \psfig{figure=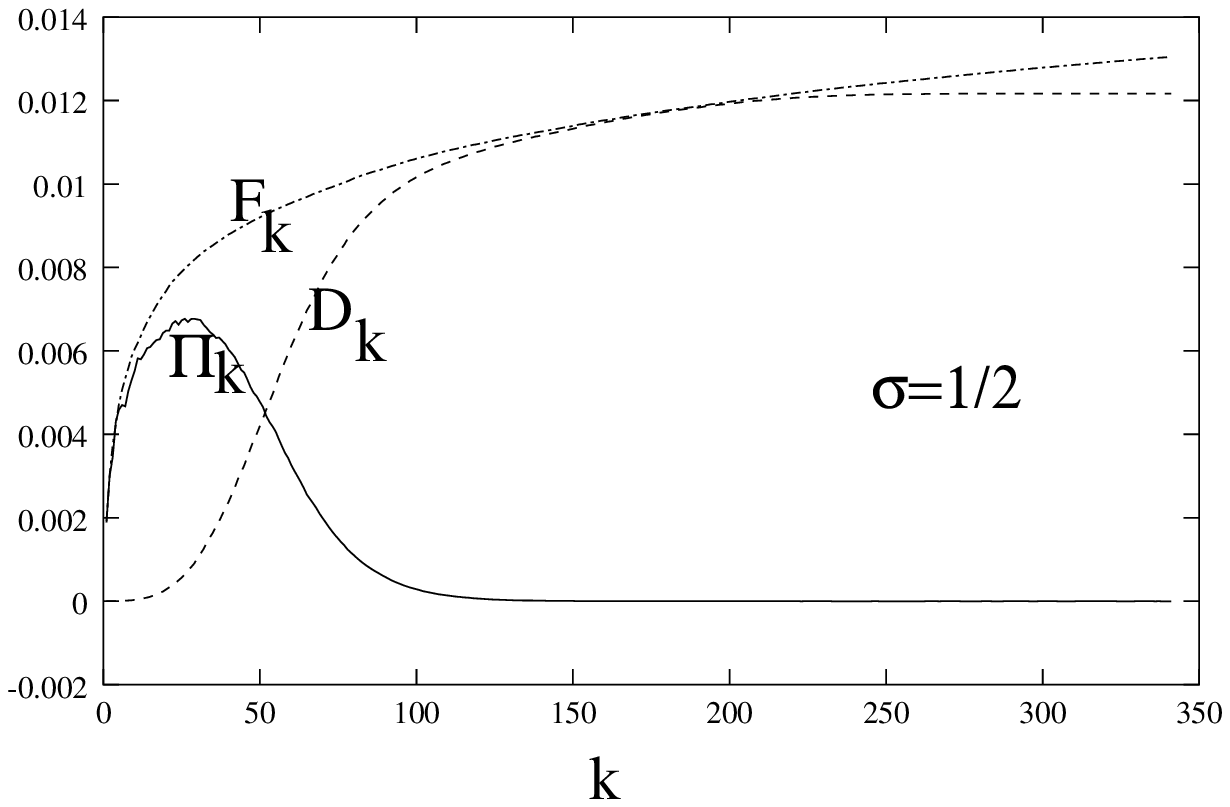,width=12cm,angle=0}}
        \vspace*{0.5cm}
\caption{}
\label{flux0.5fig}
\end{figure}

\begin{figure}
\centerline{
        \psfig{figure=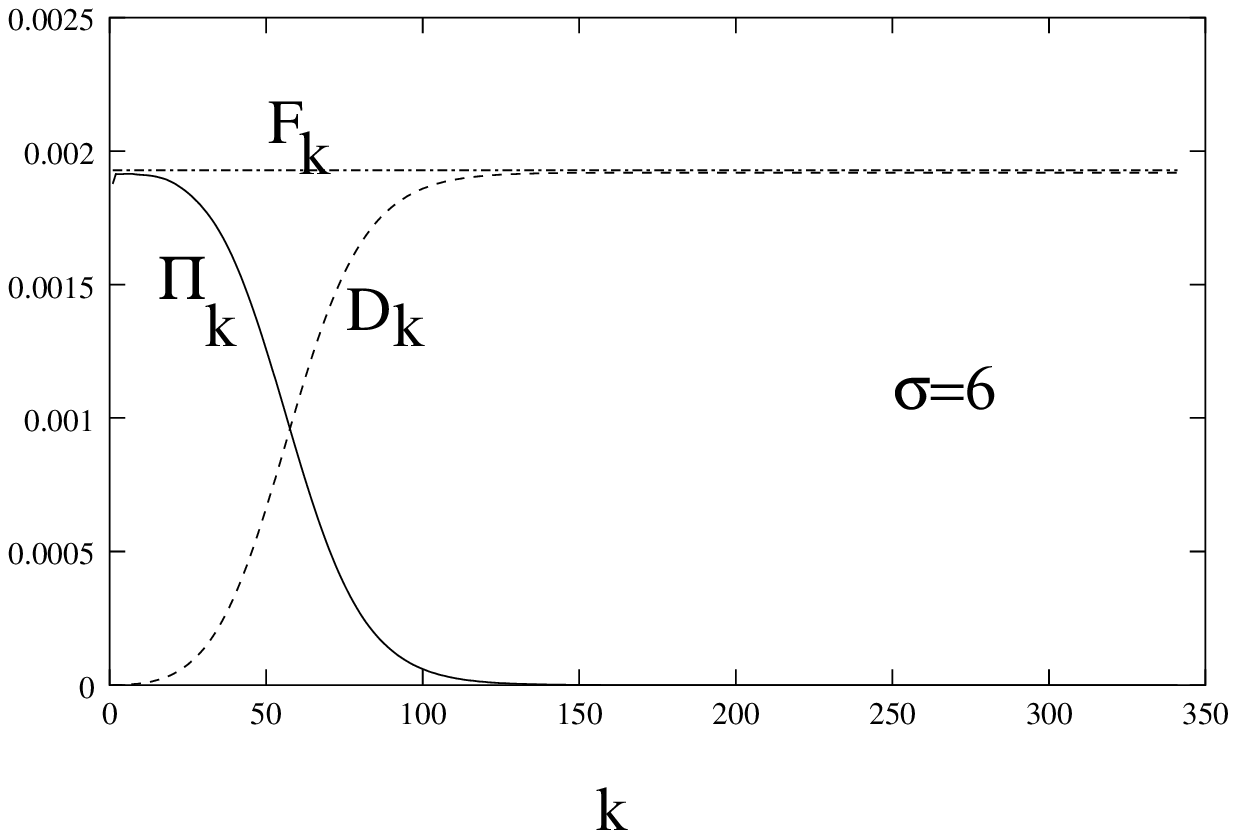,width=12cm,angle=0}}
        \vspace*{0.5cm}
\caption{}
\label{flux6fig}
\end{figure}

Fig.~\ref{flux0.5fig} shows the plots of $D_{K}, \Pi_{K}$, and $F_{K}$ for
$\sigma=1/2$.  The forcing rate $F_{K}$ is proportional to $\log(K)$.  We
also find a range of $K$ for which $\Pi_{K}$ is constant.  These
results are consistent with the findings of Chekhlov and Yakhot
\cite{Chek1}.  Fig.~\ref{flux6fig} shows the
plot of the above quantities for
$\sigma =6$.  Here again $\Pi_{K}$ is constant for a range of $K$.  An
interesting point to note is that for $\sigma=6$, $F_{K}$ is constant
beyond $K=10$ or so; this is because effectively only first few modes
are forced when $\sigma$ is large.  Similar results are obtained for all
$\sigma \ge 2$.

We find that the energy flux is constant for the noisy Burgers
equation for all $\sigma \ge 1/2$.  However, the energy spectrum varies
from $k^{-5/3}$ to $k^{-2}$ as we increase $\sigma$ from 1/2 to 3/2 and
beyond.  Hence our results show that constancy of energy flux is not a
sufficient condition for the Kolmogorov's energy spectrum in noisy
Burger equation.  This indicates that  Chekhlov and
Yakhot's \cite{Chek1} argument that the constant energy cascade rate 
for $\sigma =1/2$ implies Kolmogorov's energy spectrum is incomplete.

We find that for $\sigma \ge 2$, the flux rate $\Pi_{K} \approx 2 \times
10^{-3}$.  This numerical value is consistent with value obtained
using the formula derived by Saffman \cite{Saff} 
\be 
\Pi_{K} \approx \frac{\mu^3}{24 L}, 
\ee 
where $\mu$ is the velocity jump across the shock. In our simulations,
$\mu \approx 0.3$ for large $\sigma$.  Also, the Eq. (\ref{FK}) yields
the same numerical value ($2 D=2 \times 10^{-3}$).

\section{Discussion and Conclusions}

In this paper we have numerically calculated the energy spectrum,
structure function, and probability densities ($P(h), P(u),
P(u')$)  for the 
KPZ and Burgers equations in the presence of correlated noise.  The
Burgers equation with a strong correlated noise $(\sigma \ge 3/2)$ has
distinct shock structures.  Given this we have trivially worked out
the roughening exponents for this range: $\chi_{Burg}=1/2,
\chi_{KPZ}=3/2$.
The $\chi$ calculated from the
numerical simulation matches quite well with this result.
It has been theoretically  shown by many researchers that  
$\zeta^{Burg}_q \approx 1$
when  shocks are present. Numerically we find the same  $\zeta^{Burg}_q$ for
$\sigma \ge 3/2$.   The probability density 
$P(u') \propto u'^{-\alpha}$ with $\alpha=2.5 \pm 0.3$ for negative
$u'$; this result is in agreement with the Polyakov's theoretical
predictions.  However, for positive $u'$, our results do not match
with Polyakov's predictions for this range of $\sigma$.
Note that our results differ from those of 
E and Eijnden \cite{E} and Kraichnan \cite{KraiBurg} who obtained
$\alpha=7/2$, and Gotoh and Kraichnan \cite{GotoKrai} who argue for
$\alpha =3$

In the  $\sigma$ regime  $-1 \le
\sigma \le 0$, our numerical $\chi_{KPZ}$ matches with the RG predictions.
The exponent $\beta_{KPZ}$, however, saturates at 0.5 contrary to
the RG predictions. 
For $0 \le \sigma \le 3/2$, the exponent $\chi_{KPZ}$ varies smoothly
from 1 to 3/2.   Boldyrev's predictions appear to match with
our numerical results for $0 \le \sigma \le 1/2$.
After that there is a significant deviation. Our claims, however,
are not on a very strong ground because of 
large uncertainties, and hence, more exhaustive
 simulations are needed to reach definite conclusions. 

We find in our simulation that $P(u)$ is gaussian for all $\sigma$s and
also $\zeta^{KPZ}_q \propto q$.  Therefore, according to the convention 
discussed in the introduction, the signal $h$ will be termed
as non-intermittent for all $\sigma$s.   Note, however, that according
to the same convention, $u$ is
as intermittent for $\sigma \ge 1/2$.  These two statements  appear to be
 contradictory because $u$ and $h$
are related by $u=-\nabla h$. The apparent contradiction is quite
simple to resolve.   The velocity signal $u$, being the derivative
of $h$, is much more singular than $h$ because of the presence of the
shocks.  That is why probability density $P(u')$ is nongaussian (intermittent
$u$), even though $P(u)$ is nongaussian (non-intermittent $h$).  
This observation cautions us to choose a right variable while
investigating the system for intermittency.  Note that for fluid
turbulence also, $P(u)$ is gaussian, but $P(u')$ is not, and the
variables used for structure function is $u$ (there is no
corresponding $h$ anyway).

The nonexistence of intermittency in KPZ equation will have relevance
to other models of surface growth.  Sneppen and Jensen \cite{SnepJens}
and Tang and Leschhorn \cite{Tang} calculated the structure function
in surface growth equation in presence of quenched disorder. They find
no spatial multiscaling in their system; this result could be related
to the conclusions described in our paper.   Sneppen and
Jensen \cite{SnepJens}, and Tang and Leschhorn \cite{Tang} find
temporal intermittency in their model.  We believe temporal structure
function for the noisy KPZ and Burgers equation are important and will
shed further insights into the dynamics of these systems.

Krug \cite{Krug} and Kundagrami et
al.~\cite{Chan} have investigated the existence of intermittency in
the surface growth model of 
Das Sarma and Tamborenea (DT) \cite{DT} model and  found
multiscaling in it.  Krug, however, finds absence of
multiscaling in a variation of the DT model that was tilt independent.
The DT model is related to linear Lengevin equation which is not
expected to show intermittency, but Krug has attempted to relate
intermittency in the DT model with the existence of relevant variables of
renormalization groups. The connection of Krug's result to KPZ
equation is not clear to us at this stage.  Also, it is not clear
whether the multiscaling in the DT model is an artifact of lattice
effects, or it is due to appearance of relevant of nonlinearities in
the continuum equation \cite{Krug}.  Another important point to note
is that $u$ of fluid turbulence should probably be mapped to $h'$, not
$h$ as done in Krug's \cite{Krug} and Kundagrami et al.'s~\cite{Chan}
papers.

In the regime  $0 \le \sigma \le 3/2$, both
structures and noise coexist.  Recently, Polyakov \cite{Poly},
Boldyrev \cite{Bold1,Bold2},  Gurarie and Migdal \cite{Gura},
E and Eijnden \cite{E}, and Kraichnan \cite{KraiBurg}
have attempted to analytically solve for the spectral exponents and 
the probability density $P(u')$ for this regime.  
Many of the recent attempts are based on methods of quantum field
theory.  Considering the important role played by the shock
structures in determination of the roughening exponents,
we believe that a calculation which incorporates both structures
and noise will be very useful for such problems.  
The roughening
exponent will get contributions both from fluctuations and embedded
structures (see Appendix B).  However, a careful analysis is 
required to isolate the individual contributions. 
Krishnamurthy and Barma \cite{Supr1}
have isolated a moving pattern in the surface growth phenomena in the
presence of quenched disorder.  Their method may be applied 
here to separate the fluctuations from the structures. 

Bouchaud et al.~\cite{Bouc} have calculated the structure function
of the Burgers equation in higher dimensions using the connection
of KPZ equation to directed polymers.  They showed that $S_q(r) \propto r$.  
In Appendix C we argue that using the shock structures of the 
higher dimensional Burgers equation, one can obtain $S_q(r) \propto r$.

In this paper we have demonstrated the usefulness of structures in
calculating the dynamical exponents of the system.  
The role of
structures in dynamics is being studied in fluid turbulence,
intermittency, self organized criticality etc. 
For example, in fluid
turbulence Hatakeyema and Kambe \cite{Hata} have used the vortex
structures to calculate the scaling exponents.  Therefore, discovery
of the connections between the structures, fluctuations, and dynamics
will yield interesting insights in the nonequilibrium phenomena around
us.

\acknowledgments 
The author thanks Mustansir Barma, Supriya Krishnamurthy, and Deepak
Dhar for discussions, references, and their kind hospitality during his
stay at TIFR, where part of this work was done.  V. Subrahmanyam's
ideas and criticisms are gratefully acknowledged.  The author also
thanks J. K. Bhattacharjee, Agha Afsar Ali,  Prabal Maiti, 
S. A. Boldyrev, and S. D. Joglekar  for
discussions at various stages, and R. K. Ghosh for providing computer
time on DEC workstation.

\appendix

\section{Effects of parameters $\nu$ and $D$ }

The roughening exponents of KPZ equation are usually stated without reference
to the range of the parameter $\nu$ and $D$. Usually it is assumed that $\nu$
is small.  This is in the same spirit as in fluid turbulence.  While
performing our simulations with correlated noise,  we, however, found
interesting changeover in the behaviour of KPZ even when $\nu$ was
small.  We describe our findings below.

As discussed in the main text, for the parameter described in
Eq. (\ref{para}), we get the exponents shown in Table 1.  However, in
one of the test runs we fixed $\nu$ at a somewhat higher value
$\nu=0.05$, and chose $D=0.001$.  For these parameters we found a
transition for the energy spectrum from $k^{-2 \rho -2}$ to $k^{-2 \rho}$ 
and finally to
$k^{-\chi_{KPZ}}$, where $\chi_{KPZ}$ is given in Table 1.  To
understand this transition, we need to look at the energy spectrum of
Edward-Wilkinson (EW) and Random deposition (RD) models.

EW equation is given by 
\begin{equation}
\label{EW}
\frac{\partial h(x,t)}{\partial t}=\nu \nabla^2 h(x,t)+f(x,t). 
\end{equation}
The above equation can be solved easily in Fourier space, which yields 
\begin{equation}
h(k,\omega )=\frac{f(k,\omega )}{i\omega -\nu k^2}. 
\end{equation}
Using 
\be
\left\langle f(k,\omega )f(k^{\prime },\omega ^{\prime })\right\rangle =
	2D \frac{k^{-2 \rho}}{i (\omega-\omega')}.
\ee
we can easily show that 
\be
\left\langle h(k,t)h^{*}(k,t)\right\rangle = k^{-2 \rho - 2} \left(
A+B \exp(-2 \nu k^{2} t \right)
\ee
Therefore, EW equation at large $t$ yields $|h(k)|^{2} \propto k^{-2
\rho -2}$.

If we substitute $\nu=0$ in the Eq. (\ref{EW}), we obtain the Random 
deposition (RD) model \cite{Bara} which is 
\begin{equation}
\frac{\partial h(x,t)}{\partial t}=f(x,t). 
\end{equation}
Using a similar procedure as above, we obtain $|h(k)|^{2} \propto
k^{-2 \rho }$ for this equation.

Now we can explain the crossover from EW to RD and then to KPZ
behaviour.  In the initial phase, the width $<h^2>$ is small, hence
the viscous term dominates both noise and nonlinear term, consequently
EW behaviour is seen.  As time progress and the width increases, the
noise term becomes dominant and RD behaviour is observed.  In this
regime the width increases linearly in time \cite{Bara}.  At a later
time when the width become large enough, the nonlinear term takes
over, and KPZ like behiour is observed.  This is the reason why the
spectral index varies from EW to RD, and finally to KPZ regime.

Our findings regarding the crossover from EW to KPZ etc. show that
we should be careful in the choice of parameters while simulating KPZ
and Burgers equations.

\section{Exponents when both structures and noise are present}

A surface profile $h(x)$ can be split into two parts:  
\be
h(x)=h_s(x)+h_f(x), 
\ee
where $h_f(x)$ denotes the fluctuation, and $h_s(x)$ denotes the the
embedded structure. By definition, $<h_f(x)>=0$ where $<.>$ denotes
the ensemble average.  Therefore, $<h(x)>=h_s(x)$.

We can show that the width squared $W^2$, which is obtained by taking the
spatial average, is 
\be
\begin{array}{c}
W^2=
\overline{\left\langle \left( h(x)-\overline{h}\right) ^2\right\rangle } \\ 
=W_s^2+W_f^2
\end{array}
\ee Hence, $W^2$ is the sum of the contributions from the fluctuating
part and the structure part. Similarly, the second-order structure
function $T_2(r)$ is the sum of two contributions, 
\be
\begin{array}{c}
T(r)=
\overline{\left\langle \left| h(x+r)-h(x)\right| ^2\right\rangle } \\ 
=T_s(r)+T_f(r).
\end{array}
\ee 
Hence, in a profile with a embedded structure, the exponent is
determined by both structure and fluctuations.

There is no structure in KPZ profile for $\sigma=-1$, hence, the
roughening exponent is totally determined the fluctuations. However,
the profile for the noiseless Burgers equation has only shock
structure, which determines the roughening exponent. In KPZ equation
with correlated noise, especially for $0 < \sigma < 3/2$, we have both
structure and fluctuations.  Therefore, we will have to do a careful
analysis combining both of them to obtain the roughening exponents for
these cases.  We expect that similar analysis have to be carried out
for the surface growth profile with quenched disorder that has both
structure and fluctuations.

\section{Structure function for Burgers equation in $d \ge 2$}

\begin{figure}
\centerline{
        \psfig{figure=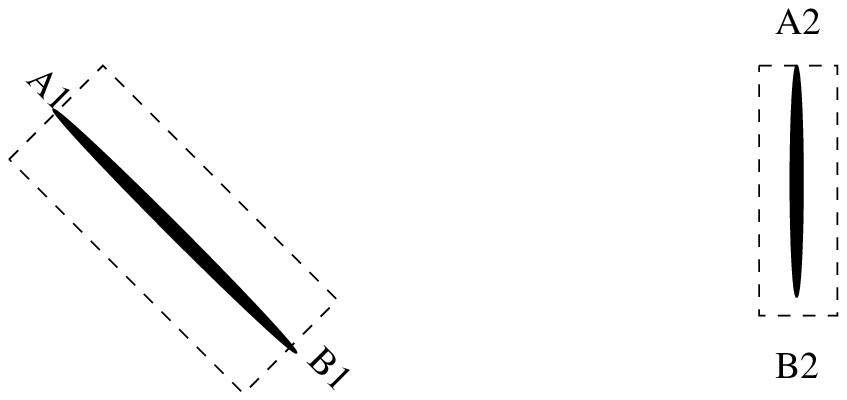,width=12cm,angle=0}}
\caption{}
\label{burgd2}
\end{figure}

The arguments of Gotoh \cite{Goto} for the computation of structure
function of the noiseless Burgers equation can be generalized to
higher dimensions. In two dimensions the projections of the shocks
will appear as shown in Fig.~\ref{burgd2}. Towards the ends (labelled
as $\ A_i$ and $B_i$ in Fig.~\ref{burgd2}) the shock strength tapers
off and goes to zero.  The velocity in between the shocks is a smooth
function, but the velocity is discontinuous at the shocks.  Therefore,
as argued in section II,
\begin{eqnarray}
  S_{q}(r)  & = & \int | {\bf u(x+r)-u(x)} |^{q} d{\bf x}      \nonumber \\
     & \approx &  C_{1} \frac{1}{A} \left(\frac{r}{t} \right)^{q} A + 
             C_{2} \frac{1}{A} \sum_{i} (\mu_{i})^{q} ( L_{i} r) 
\end{eqnarray}
where $A$ is the area of the system, $\mu _i$ and $L_i$ are the shock
strength and length of the $i$th shock respectively, and $C_1,C_2$ are
constants. The first term is obtained from the integral over all the
space except the boxed regions near the shocks, while the second term
is due to the integral in the boxed region. Essentially, we obtain
${\bf |u(x+r)-u(x)| \propto } \mu _i$ in the boxed region when ${\bf
x+r}$ is in one side of the shock, while ${\bf x}$ is on the other
side of the shock. This result yields $S_q(r)\propto r$ for $q>1$.
These arguments for two dimensions can be easily generalized to three
and higher dimensions (in three dimensions, the shock regions will
appear as disks). Hence, in any arbitrary dimensions, noiseless
Burgers equation has
\begin{equation}
S_q(r)\propto r. 
\end{equation}
Bouchaud et al. \cite{Bouc} have obtained this result in $D$
dimensions by relating to KPZ equation to 
directed polymer; their result becomes exact
as $D\rightarrow \infty $. The strong intermittency of the Burgers
equation is due to the large scale singularities of the velocity field
at the shock regions, which are concentrated in $D-1$ dimension 
(codimension 1). It was pointed out by one of the referee
that  the shocks  could be   fractal objects, whose
 fractal dimension $D_f$ could  be greater than $D-1$.
The calculation of structure function for these objects
is beyond the scope of this paper.


\centerline{\large Figure Captions}
\vspace{1cm}

\noindent
{\bf Fig. 1} \, The solutions of noiseless Burgers and KPZ equations.  \\ 

\noindent
{\bf Fig. 2} \, The probability density $P(u')$ vs. $u'/u'_{max}$ for the
noiseless Burgers equation.  \\

\noindent
{\bf Fig. 3} \, The energy spectrum $(1/2) <|h(k)|^{2}>$ vs. $k$ for
the KPZ equation when $\sigma=-1,0,1/2,2$. The lines of best fit for the
intermediate range of $k$ are also shown.  \\ 

\noindent
{\bf Fig. 4} \, The normalized velocity profile $u(x)$ vs. $x$ for Burgers
equation when $\sigma=-1,-1/2,0,1/2,1,2,6$. The profile is noisy for $-1
\le \sigma\le 0$, has only well developed shocks for $\sigma \ge 3/2$, and
has both fluctuations and shocks for the intermediate range of $\sigma$.  \\

\noindent
{\bf Fig. 5} \, The probability density $P(u)$ vs. $u/u_{max}$ for
$\sigma=-1,0,1/2$ and 6. The best fit gaussian curves are also shown in
the figure.  \\

\noindent
{\bf Fig. 6} \, The structure functions $S_{q}(r),T_q(r)$ vs. $r$
for $\sigma=6$.  \\ 

\noindent
{\bf Fig. 7} \, The exponent $\zeta^{KPZ}_q$ vs. $q$ for
$\sigma=-1,-1/2,0,1/2,2,6$.  \\ 

\noindent
{\bf Fig. 8} \, The exponent $\zeta^{Burg}_q$ vs. $q$ for
$\sigma=-1,-1/2,0,1/2,2,6$.  \\ 

\noindent
{\bf Fig. 9} \, The probability density $P(u')$ vs. $u'/u'_{max}$ for
$\sigma=-1,-1/2,0$ and 1/4. Function $0.01 \exp(-0.14 |u'|)$ fits well for
$\sigma=0$. \\

\noindent
{\bf Fig. 10} \, The probability density $P(u')$ vs. $u'/u'_{max}$ for
$\sigma=1/4,1/2,6$. The best fit curves for negative $u'$ are:
$u'^{-2.85}$ for $\sigma=1/4$, $u'^{-2.2}$ for $\sigma=1/2$, and $u'^{-2.5}$
for $\sigma=6$.  For positive $u'$, $\exp(-C x^3)$ fits well for
$\sigma=1/2$. \\

\noindent
{\bf Fig. 11} \, The Dissipation rate $D_K$, the energy flux $\Pi_K$,
and the forcing rates $F_K$ vs. K for $\sigma=0$.  \\ 

\noindent
{\bf Fig. 12} \, The Dissipation rate $D_K$, the energy flux $\Pi_K$,
and the forcing rates $F_K$ vs. K for $\sigma=1/2$.  \\ 

\noindent
{\bf Fig. 13} \, The Dissipation rate $D_K$, the energy flux $\Pi_K$,
and the forcing rates $F_K$ vs. K for $\sigma=6$. \\

\noindent
{\bf Fig. 14} \, The projection of shocks of two-dimensional Burgers
equation.  In the middle, the velocity difference across the shock is
large, while at the ends $A_{i}$ and $B_{i}$, the difference is small. 
\\

\newpage

\begin{table}
\caption{The exponents $\alpha$ and $\beta$ for KPZ equation for various 
$\sigma$'s.  The Medina et al.'s exponents are also listed for comparison.}
\begin{center} 
\begin{tabular}{lll|ll} 
\multicolumn{1}{c}{$\sigma=\rho-1$} &
\multicolumn{2}{c|}{Our results} & 
\multicolumn{2}{c}{Medina et al.'s results} \\ \hline
   & $\alpha_{KPZ}$ & $\beta_{KPZ}$ & $\alpha_{KPZ}$ & $\beta_{KPZ}$ \\ \hline
-1.00 & 0.46 & 0.35 & 0.50 & 0.33 \\
-0.85 & 0.50 & 0.28 & 0.50 & 0.33 \\
-0.75 & 0.53 & 0.31 & 0.50 & 0.33 \\
-0.60 & 0.60 & 0.33 & 0.60 & 0.43 \\
-0.50 & 0.70 & 0.36 & 0.67 & 0.50 \\
-0.25 & 0.89 & 0.42 & 0.83 & 0.71 \\
0.00 & 1.07 & 0.45 & 1.00 & 1.00 \\ \hline
0.25 & 1.15 & 0.50 &      &      \\
0.50 & 1.27 & 0.48 &      &      \\
0.75 & 1.31 & 0.50 &      &      \\
1.00 & 1.58 & 0.48 &      &      \\
2.00 & 1.54 & 0.48 &      &      \\
6.00 & 1.45 & 0.48 &      &      \\ 
\end{tabular}
\end{center}
\end{table}

\begin{table}
\caption{The structure function exponents  $\zeta^{Burg}_q$ of
$<|u(x+r)-u(x)|^q>$ for the
Burgers equation for various $\sigma$'s}
\begin{center} 
\begin{tabular}{llllllllll} 
$q \setminus \sigma=\rho-1$   & -1     & -3/4   & -1/2    & -1/4  & 0  & 1/4 & 1/2
  & 2  & 6   \\ \hline
1 & -0.012   & -0.001 & 0.017  & 0.069 & 0.16 & 0.29 & 0.42 & 0.90 & 0.94  \\
2 & -0.020   & 0.001  & 0.041  & 0.136 & 0.30 & 0.51 & 0.70 & 1.09 & 1.09  \\
3 & -0.024   & 0.006  & 0.068  & 0.197 & 0.41 & 0.63 & 0.84 & 1.00 & 1.00  \\
4 & -0.024   & 0.014  & 0.099  & 0.250 & 0.50 & 0.65 & 0.85 & 0.94 & 0.96  \\
5 & -0.020   & 0.024  & 0.130  & 0.293 & 0.55 & 0.58 & 0.81 & 0.90 & 0.93  \\
6 & -0.013   & 0.036  & 0.161  & 0.329 & 0.58 & 0.47 & 0.74 & 0.85 & 0.89  \\
7 &  0.003   & 0.052  & 0.191  & 0.358 & 0.60 & 0.34 & 0.66 & 0.81 & 0.84  \\
8 &  0.009   & 0.071  & 0.218  & 0.383 & 0.62 & 0.20 & 0.57 & 0.78 & 0.79  
\end{tabular}
\end{center}
\end{table}

\begin{table}
\caption{The structure function exponents $\zeta^{KPZ}_q$ of
$<|h(x+r)-h(x)|^q>$ for the KPZ equation for various $\sigma$'s}
\begin{center} 
\begin{tabular}{llllllllll} 
$q \setminus \sigma=\rho-1 $ & -1     & -3/4   & -1/2    & -1/4  & 0  & 1/4 & 1/2
  & 2  & 6  \\ \hline
1 & 0.45 & 0.50 & 0.62 & 0.73 & 0.82 & 0.88 & 0.92 & 0.97 & 0.97 \\
2 & 0.90 & 1.00 & 1.22 & 1.44 & 1.62 & 1.73 & 1.81 & 1.92 & 1.93 \\
3 & 1.36 & 1.49 & 1.80 & 2.12 & 2.39 & 2.55 & 2.69 & 2.86 & 2.87 \\
4 & 1.82 & 1.97 & 2.34 & 2.78 & 3.15 & 3.36 & 3.56 & 3.79 & 3.81 \\
5 & 2.27 & 2.44 & 2.86 & 3.42 & 3.90 & 4.17 & 4.42 & 4.72 & 4.75 \\
6 & 2.71 & 2.90 & 3.35 & 4.04 & 4.65 & 4.98 & 5.27 & 5.64 & 5.68 \\
7 & 3.14 & 3.38 & 3.83 & 4.65 & 5.41 & 5.78 & 6.11 & 6.55 & 6.60 \\
8 & 3.54 & 3.85 & 4.30 & 5.28 & 6.18 & 6.58 & 6.95 & 7.46 & 7.52 
\end{tabular}
\end{center}
\end{table}

\end{document}